\documentclass[acmsmall]{acmart}

\AtBeginDocument{%
  \providecommand\BibTeX{{%
    \normalfont B\kern-0.5em{\scshape i\kern-0.25em b}\kern-0.8em\TeX}}}

\usepackage[labelformat=simple]{subcaption}

\usepackage{enumitem}
\setlist{nosep}
\usepackage[skip=0.3\baselineskip]{caption}
\usepackage{bm}
\usepackage{algorithmicx,algorithm}
\usepackage{algpseudocode}
\usepackage[switch]{lineno}

\usepackage{booktabs}
\usepackage{float}
\usepackage{overpic}
\usepackage{float}  
\usepackage{subfloat}
\usepackage{bm}
\usepackage{multirow}
\usepackage{graphicx}
\usepackage{subcaption}
\usepackage{colortbl}
\usepackage{enumitem}
\usepackage{stfloats} 
\usepackage{url}
\usepackage[normalem]{ulem}
\useunder{\uline}{\ul}{}
\usepackage[skip=0.5\baselineskip]{caption}
\usepackage{algorithm,algpseudocode}
\definecolor{editBlue}{RGB}{134,150,167}

\usepackage{xspace}
\newcommand{\etal}{\emph{et al.}\xspace}
\newcommand{\eg}{\emph{e.g.,}\xspace}
\newcommand{\ie}{\emph{i.e.,}\xspace}

\setcopyright{acmcopyright}
\copyrightyear{2024}
\acmYear{2024}
\acmDOI{XXXXXXX.XXXXXXX}

\acmConference[XX]{Make sure to enter the correct
  conference title from your rights confirmation emai}
  {June 03--05,2018}{Woodstock, NY}
\acmPrice{15.00}
\acmISBN{978-1-4503-XXXX-X/18/06}

\begin{document}

\title{A Contrastive Pretrain Model with Prompt Tuning for \\ Multi-center Medication Recommendation}

\author{Qidong Liu}
\email{liuqidong@stu.xjtu.edu.cn}
\orcid{0000-0002-0751-2602}
\affiliation{%
  \institution{Xi'an Jiaotong University \& City University of Hong Kong}
  \city{Xi'an}
  \country{China}
}

\author{Zhaopeng Qiu}
\orcid{0009-0006-7067-3763}
\affiliation{%
  \institution{Jarvis Research Center, Tencent YouTu Lab}
  \city{Beijing} 
  \country{China}}
\email{zhpengqiu@gmail.com}

\author{Xiangyu Zhao}
\orcid{0000-0003-2926-4416}
\authornote{Corresponding Authors.}
\affiliation{%
  \institution{City University of Hong Kong}
  \city{Hong Kong}
  \country{China}
}
\email{xianzhao@cityu.edu.hk}

\author{Xian Wu}
\orcid{0000-0003-1118-9710}
\affiliation{%
 \institution{Jarvis Research Center, Tencent YouTu Lab}
 \city{Beijing}
 \country{China}}
\email{kevinxwu@tencent.com}
\authornotemark[1]

\author{Zijian Zhang}
\orcid{0000-0003-1194-8334}
\affiliation{%
  \institution{Jilin University \& City University of Hong Kong}
  \city{Changchun}
  \country{China}}
\email{zhangzijian@jlu.edu.cn}

\author{Tong Xu}
\orcid{0000-0003-4246-5386}
\affiliation{%
  \institution{University of Science and Technology of China}
  \city{Hefei}
  \country{China}}
\email{tongxu@ustc.edu.cn}

\author{Feng Tian}
\orcid{0000-0001-7888-0587}
\affiliation{%
  \institution{Xi'an Jiaotong Univeristy}
  \city{Xi'an}
  \country{China}}
\email{fengtian@mail.xjtu.edu.cn}
\authornotemark[1]

\renewcommand{\shortauthors}{Qidong Liu~\etal}

\begin{abstract}

    Medication recommendation is one of the most critical health-related applications, which has attracted extensive research interest recently. Most existing works focus on a single hospital with abundant medical data. However, many small hospitals only have a few records, which hinders applying existing medication recommendation works to the real world. Thus, we seek to explore a more practical setting, \ie multi-center medication recommendation. In this setting, most hospitals have few records, but the total number of records is large. Though small hospitals may benefit from total affluent records, it is also faced with the challenge that the data distributions between various hospitals are much different. In this work, we introduce a novel con\textbf{T}rastive pr\textbf{E}train \textbf{M}odel with \textbf{P}rompt \textbf{T}uning (\textbf{TEMPT}) for multi-center medication recommendation, which includes two stages of pretraining and finetuning. We first design two self-supervised tasks for the pretraining stage to learn general medical knowledge. They are mask prediction and contrastive tasks, which extract the intra- and inter-relationships of input diagnosis and procedures. Furthermore, we devise a novel prompt tuning method to capture the specific information of each hospital rather than adopting the common finetuning. On the one hand, the proposed prompt tuning can better learn the heterogeneity of each hospital to fit various distributions. On the other hand, it can also relieve the catastrophic forgetting problem of finetuning. To validate the proposed model, we conduct extensive experiments on the public eICU, a multi-center medical dataset. The experimental results illustrate the effectiveness of our model. The implementation code is available to ease the reproducibility\footnote{\url{https://github.com/Applied-Machine-Learning-Lab/TEMPT}}.

\end{abstract}



\received{xx xxxx}
\received[revised]{xx xxxx}
\received[accepted]{xx xxxx}

\maketitle

\section{Introduction} \label{sec_intro}



\begin{figure}[!t]
\centering
    \begin{minipage}[b]{0.495\linewidth}
		\centering
        \begin{subfigure}{1\linewidth}
		\includegraphics[width=0.8\linewidth]{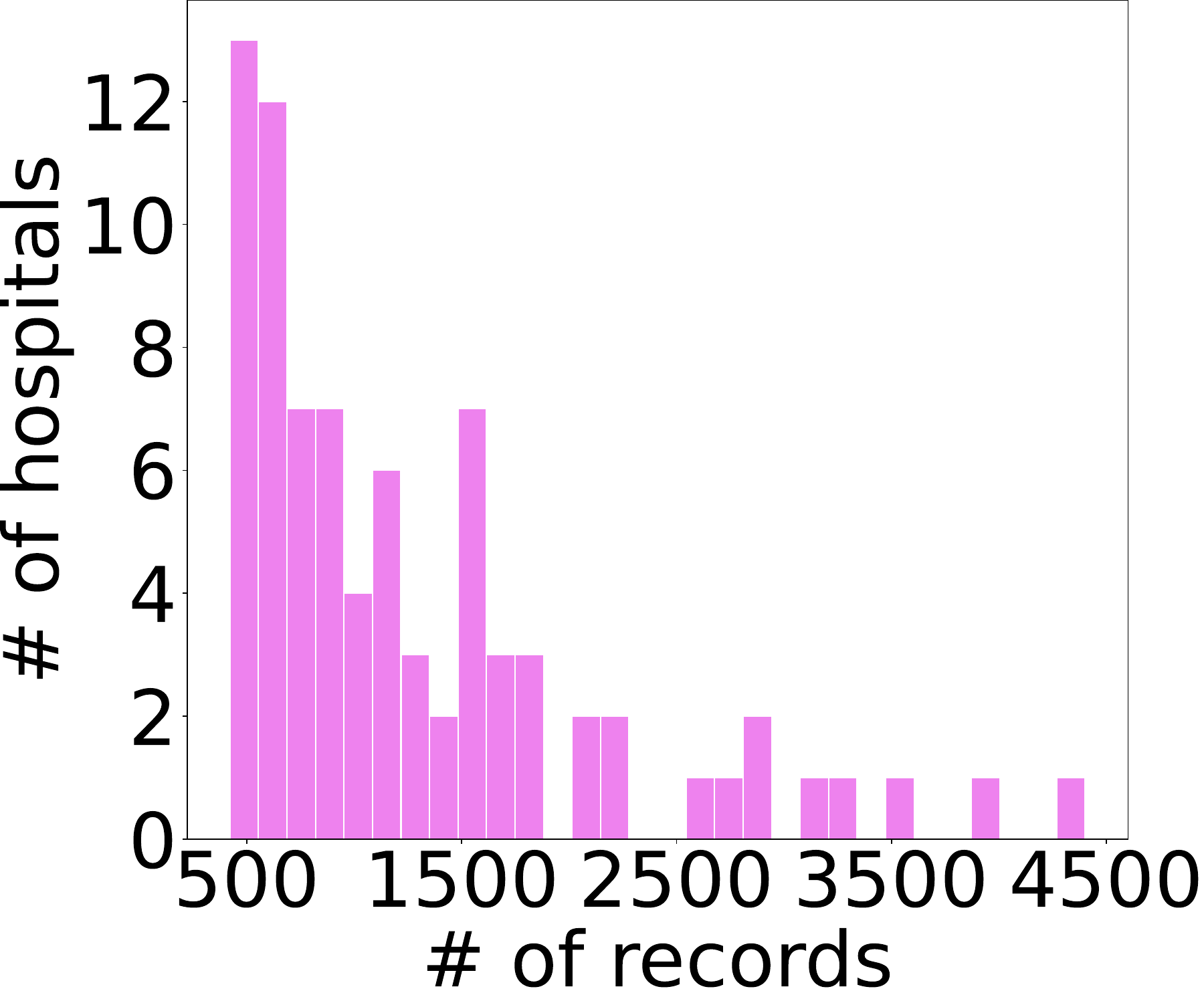}
        \caption{Record counts.}
		\label{fig_pre_reco}
        \end{subfigure}
	\end{minipage}
    \begin{minipage}[b]{0.495\linewidth}
		\centering
        \begin{subfigure}{1\linewidth}
		\includegraphics[width=0.8\linewidth]{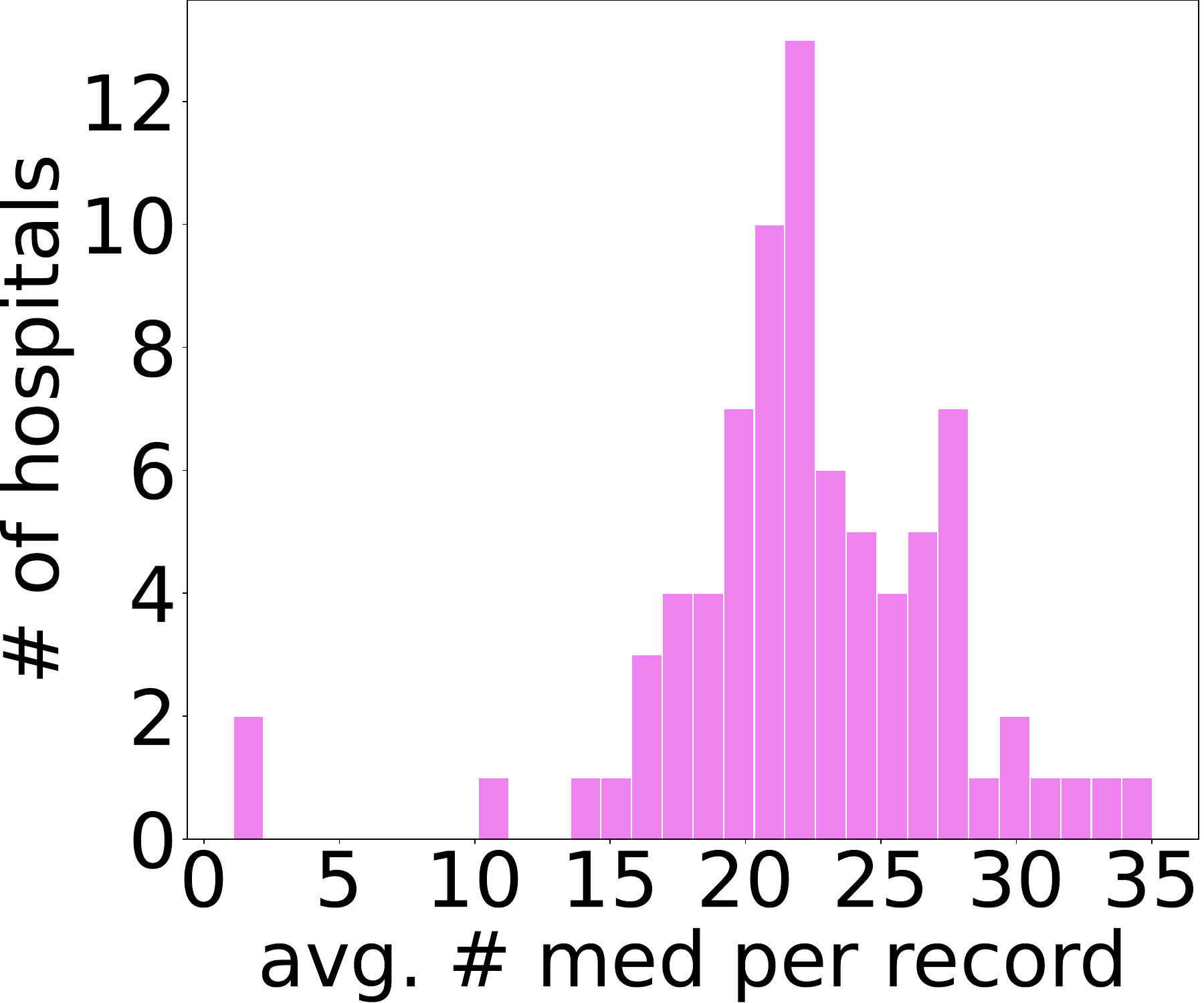}
        \caption{Average medication number.}
		\label{fig_pre_med}
        \end{subfigure}
	\end{minipage}
\caption{The histogram of medical record counts of each hospital and the histogram of the average number of prescribed medications by each hospital.}
\label{fig_pre}
\end{figure}

Prescription is an important way for doctors to treat patients, but facing patients with various diagnoses requires tremendous and sophisticated expert efforts. 
Due to high expertise and complexity, some medical research studies have reported that senior doctors have suffered from heavy workloads of prescriptions~\cite{wen2016workload,lewis2009uncomfortable}, while junior doctors are prone to error prescriptions~\cite{reynolds2017improving,barber2003reducing}. The automatic medication recommender system could potentially assist doctors and alleviate these problems~\cite{ali2023deep,mai2023drug}.
Thus, medication recommendation has drawn increasing research attention in recent years \cite{shang2019gamenet, wu2022conditional, Yang2021ChangeMM}. Thanks to the advancement of deep learning, some researchers have explored recommending proper medication sets for a single hospital with abundant medical records.
However, in the real world, the number of health records in one hospital is often too few. As shown in Figure~\ref{fig_pre_reco}, we analyze the statistics of the eICU~\cite{pollard2018eicu}, which is a real-world multi-center medical dataset. The histogram of record counts of each hospital illustrates that only a few hospitals have more than 2,500 records and most hospitals even own less than 1,500 records. Such few records are insufficient for adequately training a model, but the affluent sum of data from all hospitals gives us a chance to enhance the model. 
Therefore, we focus on making use of records from several hospitals (\ie multi-center) and constructing the model that can recommend proper medication sets for all of these hospitals, especially for benefiting those hospitals with few records.




Similar to the process of prescription in the real world, the medication recommendation models give out the proper medication set according to the patient's diagnosis and procedure. The diagnosis often contains the patient's disease, such as ``heart disease'', which should be targeted by recommended medications. As for the procedures, such as ``surgery'', they need medications for assistance to treat patients, so medications also can be indicated by them.
With the growing research interests in this field, numerous deep learning-based models have emerged recently~\cite{shang2019gamenet, wu2022conditional}, which can be divided into two categories: \textit{Longitudinal} models and \textit{Instance-based} models. The longitudinal models~\cite{shang2019gamenet,shang2019pre,Yang2021ChangeMM,yang2021safedrug,wu2022conditional} recommend medication combinations based on patient's historical records. 
However, most models in this thread cannot recommend for the patients without historical data, which is more practical in real-world scenarios \cite{tan20224sdrug}. In contrast, the instance-based models are based on the current diagnosis and procedures of one patient, and there is no need for historical information, such as Leap~\cite{zhang2017leap} and SMR~\cite{gong2021smr}. To be more practical, we will focus on the instance-based setting in this paper.

\begin{figure}[!t]
\centering
\includegraphics[width=0.8\linewidth]{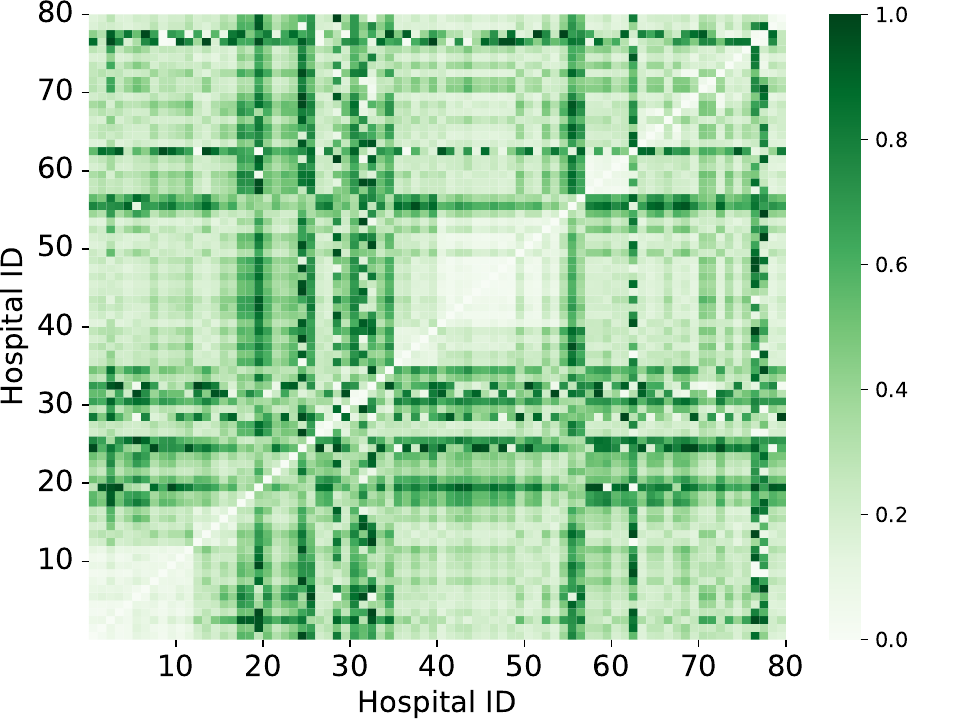}
\caption{The heatmap to visualize Jensen–Shannon divergence of prescription distribution for all pairs of hospitals in the eICU dataset.}
\label{fig_pre_dis}
\end{figure}

Existing medication recommendation models are designed for a single hospital \cite{shang2019gamenet,shang2019pre,wu2022conditional}, but often ignore the multi-center setting.
At the same time, the variance of access to treatment resources among multiple hospitals may cause distinct therapies and prescriptions. For example, some research studies have explored how different hospital conditions affect patient outcomes, such as the bronchoscopy~\cite{wayne2022variation} and bed size~\cite{shahin2019connected}. Besides, whether a hospital is a teaching hospital also affects the ways of treatment~\cite{rajabalizadeh2020exploratory}. These previous research papers indicate the existence of prescription distribution variance, which may hinder unifying a model for multi-center medication recommendation. To illustrate the distribution variance further, we demonstrate the results of two preliminary experiments.
As shown in Figure~\ref{fig_pre_med}, the average prescribed medication number varies much across hospitals, which illustrates that the differences between distributions of various hospitals are large.
Besides, we show the Jesen-Shannon divergence of prescription distributions for all pairs of hospitals in the eICU dataset. Jesen-Shannon divergence ($JSD$)~\cite{lin1991divergence} is an entropy-based method to measure the similarity between two distributions. In specific, $JSD$ is bounded between $0$ and $1$, where $JSD=0$ means the same two distributions and vice versa. In this experiment, we calculate the $JSD$ of prescription distributions between all pairs of hospitals. As shown in Figure~\ref{fig_pre_dis}, the heatmap indicates that the $JSD$ values between most hospital pairs are larger than $0.4$ and some even reach above $0.8$. The results further demonstrate the great prescription distinction between hospitals.
Such a difference will degrade the model trained on multi-center data. 
Recently, some efforts on multi-domain recommendation (MDR) could relieve the problem: (i) one is multi-task learning related works, such as MMOE~\cite{ma2018modeling} and PLE~\cite{tang2020progressive}. However, they often work for a small number of domains but are not fit for too many centers, because they need the same number of expert networks as the domains~\cite{jiang2022adaptive}; (ii) the other is about multi-domain Click Through Rate (CTR) prediction, \eg STAR~\cite{sheng2021one}, which devises a star topology learning scheme. The shortcoming of this work is that it is built on a feed-forward network and cannot combine with advanced architectures used in medication recommendation, such as transformer~\cite{wu2022conditional}.


For the challenge of multi-center medication recommendation mentioned above, we propose a novel con\textbf{T}rastive pr\textbf{E}train \textbf{M}odel with \textbf{P}rompt \textbf{T}uning (\textbf{TEMPT}) for multi-center medication recommendation. To extract the medical information from diagnosis and procedures, we construct a transformer-based recommendation model. Specially, this model consists of two stages: pretraining and prompt tuning. At the pretraining stage, we propose two self-supervised tasks to learn general medical knowledge from all hospitals. The diagnosis and procedure contain medical information in themselves, so we design a mask prediction task that reconstructs them to learn the latent intra-relationships. Then, due to the close inter-relationship between diagnosis and procedure sets, we propose a novel contrastive task to capture such many-to-many relationships via aligning the pair of representations. After pretraining, finetuning for each hospital is an instinctive way. However, the general finetuning method faces the problem of catastrophic forgetting~\cite{ramasesh2022effect}. 
Inspired by the advancement of prompt learning in NLP~\cite{li2021prefix} and CV~\cite{jia2022visual}, we design a continuous prompt tuning method for the pretrained model, which can both hold the knowledge in the pretraining stage and better extract specific information for each hospital.
The main contributions of this paper can be concluded as follows:
\begin{itemize}[leftmargin=*]
    \item To the best of our knowledge, this paper is the first to investigate the multi-center setting in the medication recommendation field. This setting is closer to the situation of the real world, where small hospitals occupy the majority. 
    
    \item We propose a two-stage model for multi-center medication recommendation. At the pretraining stage, we devise mask prediction and contrastive self-supervised tasks to learn general medical knowledge. Then, we design an efficient prompt-tuning method to adapt to the distribution of each hospital.
    
    \item Extensive experiments on a multi-center medical dataset eICU have been conducted. The experimental results demonstrate that the proposed model outperforms state-of-the-art medication recommendation models and multi-domain recommendation models.
    
\end{itemize}

The rest of this paper is organized into five sections. In Section~\ref{sec_prob}, we formulate the multi-center medication recommendation. Then, we detail the proposed model in Section~\ref{sec_model}. The experiments and analysis are given in Section~\ref{sec_exp}. Finally, we refer to the related works and conclusion in Section~\ref{sec_related} and Section~\ref{sec_conclusion}, respectively.

\section{Problem Formulation} \label{sec_prob}

\begin{table}[]
    \centering
    \caption{Notations used in TEMPT.}
    \resizebox{0.9\linewidth}{!}{%
    \begin{tabular}{ll}
    \toprule[1pt]
    \textbf{Notation} & \textbf{Description} \\ 
    \midrule
    \midrule
    $\mathcal{V}$  &  Records of all hospitals      \\
    $\mathcal{R}^{(h)}$  &  Records of the $h$-th hospital     \\  \hline
    $\mathcal{D}_U$, $\mathcal{P}_U$, $\mathcal{M}_U$    &   The set of diagnosis, procedures and medication     \\ 
    $\mathbf{E}_d$, $\mathbf{E}_p$  &  Embedding matrices for diagnosis and procedures        \\
    $\mathbf{e}_k$  &  The embedding of mask token \\
    $\mathbf{D}$, $\mathbf{P}$   &  Input embedding set of original diagnosis and procedures     \\
    $\mathbf{D}'$, $\mathbf{P}'$   &  Input embedding set of masked diagnosis and procedures     \\
    $\mathbf{D}''$, $\mathbf{P}''$   &  Input embedding set of prompted diagnosis and procedures     \\
    $\mathbf{r}_d$, $\mathbf{r}_p$  &  The medical representation of original diagnosis and procedure set     \\
    $\mathbf{r}_d'$, $\mathbf{r}_p'$  &  The medical representation of masked diagnosis and procedure set     \\
    $\mathbf{r}_d''$, $\mathbf{r}_p''$  &  The medical representation of prompted diagnosis and procedure set     \\
    $\mathbf{E}_h^d$, $\mathbf{E}_h^p$  &  Prompt embedding matrices of diagnosis and medication \\
    $\mathbf{o}_d$, $\mathbf{o}_p$  &   The embedding of prompt for diagnosis and procedure inputs  \\
    $\Theta_{encoder}$  &   The parameters of the medical encoder \\
    $\Theta_{MLP}^{(h)}$  &     The parameters of ${\rm MLP}_r$ for hospital $h$
    \\ \hline
    $\mathbf{y}_d$  &  Labels of masked diagnosis \\
    $\mathbf{y}_p$  &  Labels of masked procedures \\
    $\mathbf{y}$   &   Labels of recommended medications  \\
    \bottomrule[1pt]
    \end{tabular}
    \label{tab_notation}}
\end{table}

The multi-center Electrical Health Records (EHR) consist of uniform data from several hospitals. Let $\mathcal{V}=\{\mathcal{R}^{(h)}\}^H_{h=1}$ denote the integrated EHR with $H$ hospitals. In the EHR of each hospital, the records can be represented as $\mathcal{R}^{(h)}=[\mathcal{R}^{(h)}_1,...,\mathcal{R}^{(h)}_s,...,\mathcal{R}^{(h)}_{S^{(h)}}]$, where $S^{(h)}$ is the number of records of hospital $h$. For simplicity, we omit the hospital index $h$ for the single record when we describe the proposed model. We let $\mathcal{D}_U=\{d_1,d_2,...,d_{|\mathcal{D}_U|}\}$ denote the diagnosis set, $\mathcal{P}_U=\{p_1,p_2,...,p_{|\mathcal{P}_U|}\}$ denote the procedure set, and $\mathcal{M}_U=\{m_1,m_2,...,m_{|\mathcal{M}_U|}\}$ denote the medication set, where $|\mathcal{D}|$, $|\mathcal{P}|$ and $|\mathcal{M}|$ represent the number of all possible diagnosis, procedures and medications, respectively. Each record includes three types of set, \ie $\mathcal{R}_s=\{\mathcal{D},\mathcal{P},\mathcal{M}\}$, where $\mathcal{D} \in \mathcal{D}_U$, $\mathcal{P} \in \mathcal{P}_U$ and $\mathcal{M} \in \mathcal{M}_U$. Based on the notations mentioned above, we can formulate the problem of multi-center medication recommendation as: 
\textit{given a patient from any hospital $\mathcal{R}^{(h)}$ with $\mathcal{D}$ and $\mathcal{P}$, we aim to train a model on the multi-center EHR $\mathcal{V}$, which can recommend proper medication set $\mathcal{M}$ for the user.} The important notations used in this paper are listed in Table~\ref{tab_notation}.

\section{The Proposed Model} \label{sec_model}

In this section, we illustrate the overview of TEMPT first. Then, the input representations and medical encoder are introduced. Finally, we detail the procedure of pretraining and prompt tuning.

    \subsection{Overview}

        The overview of the proposed TEMPT consists of two stages: \textit{pretraining} and \textit{prompt tuning}. For recommending suitable medication combinations, we first use the embedding matrix to get a dense representation of each diagnosis and procedure. Then, two stacked transformer layers, named the medical encoder, are used to encode the set of diagnosis and procedure embeddings. To learn the general knowledge from multi-center, the model is trained on $\mathcal{V}$ during the pretraining stage, which is illustrated in Section~\ref{sec_Pretraining}. Then, we conduct prompt tuning on EHR $\mathcal{R}^{(h)}$ to adapt to each hospital at the second stage, which is detailed in Section~\ref{sec_PromptTuning}.

    
    \subsection{Input Representation and Medical Encoder}

    In general, the patients have the records of several diagnosis and procedures simultaneously, which causes complicated combinations of the input. To extract the meticulous medical information of patients, we design the input representation and medical encoder.

        \subsubsection{Input Representation} We assign two learnable embedding matrices to convert the diagnosis and procedure codes into dense vectors. Each row of the matrices refers to one unique code of diagnosis or procedure. Let $\mathbf{E}_d \in \mathbb{R}^{|\mathcal{D}_U|\times c}$ denote the embedding matrix for diagnosis and $\mathbf{E}_p \in \mathbb{R}^{|\mathcal{P}_U|\times c}$ for the procedure, where $c$ represents the embedding dimension. Then, for the diagnosis and procedures set, we can get the embedding set $\mathbf{D}=[\mathbf{e}_d^1,...,\mathbf{e}_d^i,...,\mathbf{e}_d^N]$ for diagnosis and the embedding set $\mathbf{P}=[\mathbf{e}_p^1,...,\mathbf{e}_p^j,...,\mathbf{e}_p^M]$ for procedure, respectively, via the two embedding matrices, where $N$ and $M$ are the numbers of diagnosis and procedure in the input sets. The two embedding sets are denoted as input representations and will be input to the medical encoder.

        \subsubsection{Medical Encoder} The medical encoder consists of two transformer architectures~\cite{vaswani2017attention} to encode the patients' health conditions from the aspect of diagnosis and procedures. To make use of similar medical knowledge of diagnosis and procedure during the two-stage training, two architectures share the same parameters. Each architecture has $K$ transformer layers. We denote the parameters of the medical encoder as $\Theta_{encoder}$.
        
        To capture the relations hidden in the input diagnosis and procedure sets, the transformer layer has two crucial sequential components, \ie multi-head attention and point-wise feed-forward network. We will take the input diagnosis representation as an example to illustrate it. In multi-head attention, the attention function is the basic unit, which can be defined as follows:
        \begin{equation} \label{eq1}
            {\rm Attention}(\textbf{Q},\textbf{K},\textbf{V})={\rm Softmax}(\frac{\textbf{Q} \textbf{K}^\textbf{T}}{\sqrt{c}})\textbf{V} 
        \end{equation}

        \noindent where $\textbf{Q}\in \mathbb{R}^{L_Q \times c}$, $\textbf{K}\in \mathbb{R}^{L_K \times c}$ and $\textbf{V}\in \mathbb{R}^{L_V \times c}$ are the input matrices. $c$ is the dimension of matrices. Then, several identical attentions compose the multi-head attention to obtain different views of the medical information:
        \begin{equation} \label{eq2}
            \begin{aligned}
                {\rm MultiHead}(\textbf{Q},\textbf{K},\textbf{V})&=[{\rm head}_1||...||{\rm head_a}||...||{\rm head_A}]\textbf{W}^O \\
                {\rm head}_a&={\rm Attention}(\textbf{Q} \textbf{W}_a^Q, \textbf{K} \textbf{W}_a^K, \textbf{V} \textbf{W}_a^V)
            \end{aligned}
        \end{equation}

        \noindent where $\textbf{W}_a^Q \in \mathbb{R}^{c\times (c/A)}$, $\textbf{W}_a^K \in \mathbb{R}^{c\times (c/A)}$, $\textbf{W}_a^V \in \mathbb{R}^{c\times (c/A)}$ and $\textbf{W}^O \in \mathbb{R}^{c\times c}$ are all the weight matrices. $A$ represents the number of heads. $||$ is the operation of concatenation. 
        With the residual connection, the medium output $\mathbf{M}$ from the multi-head attention can be represented as:
        \begin{equation} \label{eq3}
            \mathbf{M}={\rm LayerNorm}(\mathbf{D}, {\rm MultiHead}(\mathbf{D},\mathbf{D},\mathbf{D}))
        \end{equation}

        Next, a feed-forward network with residual connection is imposed on the medium output $\mathbf{M}$ and we can get the output of the transformer layer as:
        \begin{equation} \label{eq4}
            \begin{aligned}
                {\rm FFN}(\mathbf{M})&={\rm ReLU}(\textbf{M} \textbf{W}_1^F+\textbf{b}_1^F)\textbf{W}_2^F+\textbf{b}_2^F \\
                \mathbf{D}^{(1)}&={\rm LayerNorm}(\mathbf{M}, {\rm FNN}(\textbf{M}))
            \end{aligned}
        \end{equation}

        \noindent where $\textbf{W}_1^F\in \mathbb{R}^{c \times c}$, $\textbf{W}_2^F\in \mathbb{R}^{c \times c}$, $\textbf{b}_1^F\in \mathbb{R}^{c}$ and $\textbf{b}_2^F\in \mathbb{R}^{c}$ are trainable parameters. ${\rm LayerNorm}(\cdot)$ represents the standard layer normalization function. The $\mathbf{D}^{(1)}$ is the output of the first transformer layer for the diagnosis or procedure representation. 
        
        Similarly, we can get the set of diagnosis and procedure representation vectors from $K$ transformer layers and they are denoted as $\mathbf{D}^{(K)}$ and $\mathbf{P}^{(K)}$. To get the comprehensive representation of the whole diagnosis and procedure sets, we only take out the first row of embedding of $\mathbf{D}^{(K)}$ and $
        \mathbf{P}^{(K)}$ as the medical representation of diagnosis and procedure and we let $\mathbf{r}_d$ and $\mathbf{r}_p$ denote them, respectively. Note that we insert ``[CLS]'' token to the first position of diagnosis and procedure set as common BERT~\cite{devlin2018bert} does, so $\mathbf{r}_d$ and $\mathbf{r}_p$ are actually corresponding to the ``[CLS]'' token in the transformer. The transformer embeds the information of the whole sequence in such a special token. 
        Another noteworthy point is that we do not apply the positional embedding as the original transformer, because the elements in diagnosis and procedure sets are not strictly ordered.

        The outputs of the medical encoder, \ie $\mathbf{r}_d$ and $\mathbf{r}_p$, represent the extracted information of the patient's diagnosis and procedure, respectively. Then, they are fed into various tasks for pretraining and tuning, which are referred to in the next two subsections. Besides, they are also the major component for medication recommendation while inference.
    
    \subsection{Pretraining}\label{sec_Pretraining}

        \begin{figure}[!t]
            \centering
            \includegraphics[width=0.7\linewidth]{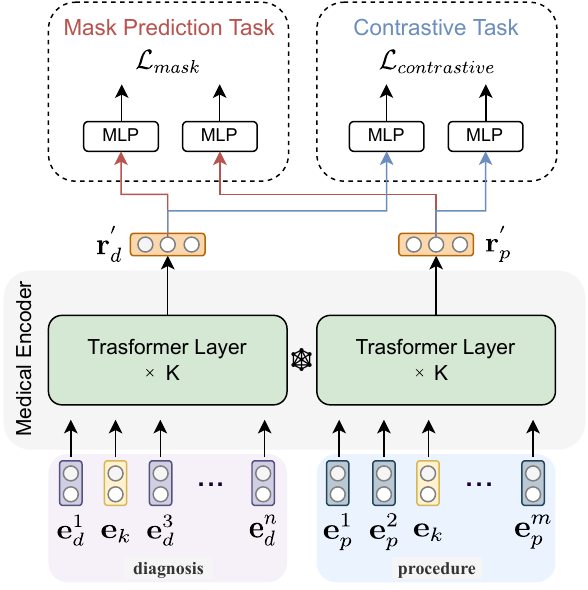}
            \caption{The pretraining stage for TEMPT.}
            \label{fig_pretrain}
        \end{figure}
    
        During the pretraining stage, we train the model on the whole multi-center records $\mathcal{V}$ to learn the general medical knowledge from all hospitals. However, it is challenging to capture such knowledge because of the data sparsity~\cite{zhou2020s3}. Besides, the correlation between diagnosis and procedures has rarely been considered before. Therefore, we design two self-supervised tasks for pretraining the model, \ie \textit{Mask Prediction Task} and \textit{Contrastive Task}, as shown in Figure~\ref{fig_pretrain}.

        \subsubsection{Mask Prediction Task} The co-occurrence relationship of medication in EHR has been proven as important information for medication recommendation~\cite{shang2019gamenet}. However, such intra-relationships between diagnosis and procedure sets are ignored at all. Inspired by the masked language model~\cite{devlin2018bert}, we devise the Mask Prediction Task for the pretraining stage to capture the co-occurrence relationships of diagnosis and procedures.

        For the mask prediction task, we randomly mask a proportion of diagnosis and procedures in these two input sets. An embedding $\mathbf{e}_k \in \mathbb{R}^c$ is assigned to the special token ``[MASK]''. Then, the input representations are converted to $\mathbf{D}'=[\mathbf{e}_d^1,\mathbf{e}_k,\mathbf{e}_d^3,...,\mathbf{e}_d^N]$ and $\mathbf{P}'=[\mathbf{e}_p^1,\mathbf{e}_p^2,\mathbf{e}_k,...,\mathbf{e}_p^M]$ as an example. After inputting $\mathbf{D}_s'$ and $\mathbf{P}_s'$ to the medical encoder, as illustrated in the last subsection, we can get the medical representation of diagnosis and procedure, which are denoted as $\mathbf{r}_d'$ and $\mathbf{r}_p'$, respectively.

        To predict the masked diagnosis and procedures, respectively, we use two multiple-layer perceptron (MLP) and sigmoid to compute the probability of each unit:
        \begin{equation} \label{eq5}
            \mathbf{\hat{y}}_d={\rm Sigmoid}({\rm MLP}_{m1}(\mathbf{r}_d'))
        \end{equation}

        \noindent where $\mathbf{\hat{y}}_d$ is the predicted masked probability of each diagnosis. Given the true label of the masked diagnosis, denoted as $\mathbf{y}_d$, we can get the loss function of the mask prediction task for diagnosis:
        \begin{equation} \label{eq6}
            \mathcal{L}_d=-\mathbf{y}_d \log(\hat{\mathbf{y}}_d) - (1-\mathbf{y}_d) \log(1-\hat{\mathbf{y}}_d)
        \end{equation}

        Similarly, $\mathbf{r}_p'$ is fed into an ${\rm MLP}_{m2}(\cdot)$ with trainable parameters and a sigmoid layer, as shown in Eq.~\eqref{eq5}, to get the predicted probability of masked procedures, which is denoted as $\mathbf{\hat{y}}_p$. Then, based on the true label of the masked procedures $\mathbf{y}_p$, we follow the Eq.~\eqref{eq6} and get the loss of the mask prediction task for procedures, denoted as $\mathcal{L}_p$. Finally, the loss of the mask prediction task is the sum of these two losses and it can be written as follows: 
        \begin{equation} \label{eq7}
            \mathcal{L}_{mask}=\sum_{\mathcal{V}}\sum_{\mathcal{R}} \mathcal{L}_d + \mathcal{L}_p
        \end{equation}

        \subsubsection{Contrastive Task} It is common that the procedures are scheduled according to the diagnosis, so they have a correspondence relation in some way. For example, the procedure Antiarrhythmic is often used to treat the diagnosis of Tachycardias. In order to capture such an inter-relationship between diagnosis and procedures, we propose a novel contrastive task for pretraining to align a pair of input diagnosis and procedures. 

        For the contrastive task, we firstly adopt two MLPs, also named projector, to project the representation $\mathbf{r}_d'$ and $\mathbf{r}_p'$ to another underlying space for a better performance \cite{chen2020simple,chen2020improved}:
        \begin{equation} \label{eq8}
            \begin{aligned}
                \mathbf{u}_d&={\rm MLP}_d(\mathbf{r}_d')  \\
                \mathbf{u}_p&={\rm MLP}_p(\mathbf{r}_p') 
            \end{aligned}
        \end{equation}

        Let $[\mathbf{u}_d^1,...,\mathbf{u}_d^i,...,\mathbf{u}_d^B]$ and $[\mathbf{u}_p^1,...,\mathbf{u}_p^j,...,\mathbf{u}_p^B]$ denote a batch of projected diagnosis and procedure representations, respectively, where $B$ is the batch size. Then, to pull the distance of paired diagnosis and procedures in one identical record, we define $\mathbf{u}_d^i$ and $\mathbf{u}_p^j$ as a positive pair when $i = j$. By comparison, any other procedure representations in this batch become negatives toward this diagnosis $\mathbf{u}_d^i$. Therefore, to push the negatives and pull the positives, we can formulate the contrastive loss function for a batch of diagnosis as follows:
        \begin{equation} \label{eq9}
            \mathcal{L}_{dp}=-\frac{1}{B} \sum_{i=1}^{B} \log \frac{\exp(sim(\mathbf{u}_d^i, \mathbf{u}_p^i) / \tau)}{\sum_{j=1}^B \mathbb{I}_{[i \neq j]} \exp(sim(\mathbf{u}_d^i, \mathbf{u}_p^j) / \tau)}
        \end{equation}

        \noindent where $\mathbb{I}_{[i \neq j]} \in \{0,1\}$ is an indicator function and $\tau$ represents the temperature parameter. $sim(\cdot, \cdot)$ is the function to evaluate the similarity between two vectors and the cosine similarity function is used in this paper.

        In $\mathcal{L}_{dp}$, the projected diagnosis representations are always in the positive pairs and contrasted with the negative procedures, so there exists another contrastive loss to push distance between the procedures and the corresponding negative diagnosis. We exchange the position between diagnosis and procedure in Eq.~\eqref{eq9}, and get the contrastive loss $\mathcal{L}_{pd}$ for the procedure. Finally, the loss function for the contrastive task can be formulated as:
        \begin{equation} \label{eq10}
            \mathcal{L}_{contrastive}=\sum_{\mathcal{V}} \sum_{\mathcal{R}} \mathcal{L}_{dp} + \mathcal{L}_{pd}
        \end{equation}

        \subsubsection{Objective} Finally, we optimize the model on all hospital records $\mathcal{V}$. The overall objective function combines the loss functions from the two self-supervised tasks mentioned above. Due to the different scales of mask prediction loss and contrastive loss, we impose a scale weight $\gamma$ when summing these two losses:
        \begin{equation} \label{eq11}
            \mathcal{L}_{pretrain}=\mathcal{L}_{mask}+\gamma \cdot \mathcal{L}_{contrastive}
        \end{equation}

        In practice, we adopt the gradient descent method to train the whole model. The learnt general medical knowledge are embedded into the parameters $\mathbf{E}_d$, $\mathbf{E}_p$ and $\Theta_{encoder}$.

        \begin{figure}[!t]
            \centering
            \includegraphics[width=0.7\linewidth]{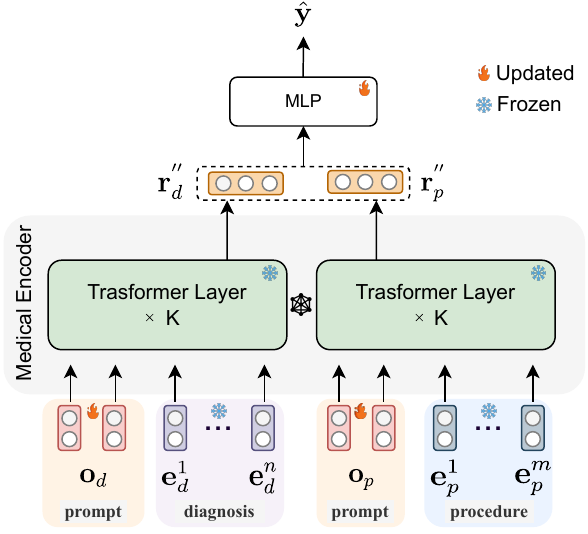}
            \caption{The prompt tuning stage for TEMPT.}
            \label{fig_prompt}
        \end{figure}

    \subsection{Prompt Tuning} 
    \label{sec_PromptTuning}
    For recommending a proper medication set, we use the pretrained diagnosis and procedure embedding matrices and medical encoder to get the medical representations of diagnosis and procedures. Then, an MLP and a Sigmoid are adopted to output the recommending probability of each medication.
         \begin{equation}   \label{eq12}
             \hat{\mathbf{y}}={\rm Sigmoid}({\rm MLP}_r([\mathbf{r}_d||\mathbf{r}_p]))
         \end{equation}

         Aimed at the problem of various hospitals with largely different distributions, finetuning the pretrained model and the ${\rm MLP}_r$ on each hospital's EHR $\mathcal{R}^{(h)}$ is an instinctive idea. The loss function can be written as follows:
         \begin{equation}   \label{eq13}
             \mathcal{L}_{tune}=-\sum_{\mathcal{R}^{(h)}} \mathbf{y} \log(\hat{\mathbf{y}}) + (1-\mathbf{y}) \log(1-\hat{\mathbf{y}})
         \end{equation}

         \noindent where $y$ is the true label of prescribed medications.

         Though the finetuning method can alleviate the problem of various distributions, it cannot capture much specific information for each hospital. Besides, finetuning also faces the challenge of catastrophic forgetting~\cite{ramasesh2022effect}, because it will modify the pretrained parameters according to the specific distribution of each hospital. Borrowing the idea of prompt tuning from natural language processing~\cite{li2021prefix} and computer vision~\cite{jia2022visual}, we propose a hospital-based prompt tuning method to tackle the problems that exist in the finetuning method, as shown in Figure~\ref{fig_prompt}.

         The process of the devised prompt tuning is detailed here. Firstly, we design prompt embeddings for learning the heterogeneous information for each hospital specifically. Let $\mathbf{E}_h^d \in \mathbb{R}^{H \times (b \times c)}$ and $\mathbf{E}_h^p \in \mathbb{R}^{H \times (b \times c)}$ denote the prompt embedding matrix for diagnosis and procedure respectively, where $b \in \{1, 2, 3...\}$ represents the number of input prompt embeddings. Each row in the two embedding matrices represents the prompt embedding for a unique hospital. We will split the $b \times c$ dimensional embedding into $b$ fractions, and each fraction is of $c$ dimension. Therefore, the prompt embedding set can be represented as $\mathbf{O}^{(h)}=[\mathbf{o}_{(h)}^1, \mathbf{o}_{(h)}^2,...,\mathbf{o}_{(h)}^b]$, \ie each hospital $h$ has its own corresponding parameters of prompt embedding.

\let\oldnl\nl
\newcommand{\nonl}{\renewcommand{\nl}{\let\nl\oldnl}}
\begin{algorithm}[!t]
\caption{The Overall Optimization Algorithm of TEMPT} \label{alg_TEMPT}
\raggedright
\textbf{Input:} The multi-center records $\mathcal{V}=\{\mathcal{R}^{(h)}\}_{h=1}^H$ \\
\textbf{Output:} The well-trained medication recommendation models for each hospital   \\
\textbf{Stage 1:} Pretraining Stage
\begin{algorithmic} [1]

\While{not converge}
\State Sample a mini-batch data instances from records of all hospitals $\mathcal{V}$
\State Random mask diagnosis and procedures for each instance
\State Calculate the loss for the mask prediction task via Eq.~\eqref{eq7}
\State Calculate the loss for the contrastive task via Eq.~\eqref{eq10}
\State Combine the loss of two tasks, take the gradient and update respective parameters $\{\mathbf{E}_d$, $\mathbf{E}_p$ and $\Theta_{encoder}\}$
\EndWhile
\end{algorithmic}

\textbf{Stage 2:} Prompt Tuning Stage
\setcounter{algorithm}{6}
\begin{algorithmic} [1]
\setcounter{ALG@line}{7}
\For{$h = 1$ to $H$}
\While{not converge}
\State Sample a mini-batch data instances from $h$-th hospital's records $\mathcal{R}^{(h)}$
\State Insert prompt embedding to the input representation
\State Calculate loss via Eq.~\eqref{eq13}
\State Take the gradient and update respective parameters $\{\mathbf{O}^{(h)}_d, \mathbf{O}^{(h)}_p, \Theta_{MLP}^{(h)}\}$
\EndWhile
\EndFor

\end{algorithmic}
\end{algorithm}

         Then, the prompt embeddings will be inserted at the start of both the input diagnosis and procedure embedding sets. For simplicity, we omit the subscript $(h)$ of prompt embeddings to illustrate. As shown in Figure~\ref{fig_prompt}, the input representations are converted to $\mathbf{D}''=[\mathbf{o}_d^1,...,\mathbf{o}_d^b, \mathbf{e}_d^1,\mathbf{e}_d^2,...,\mathbf{e}_d^N]$ and $\mathbf{P}''=[\mathbf{o}_p^1,...,\mathbf{o}_p^b,\mathbf{e}_p^1,\mathbf{e}_p^2,...,\mathbf{e}_p^M]$ after insertion. Encoded by the medical encoder and decoded by the MLP layer and sigmoid, we can get the recommending probability for each medication, which is written as follows: 
         \begin{equation}   \label{eq14}
             \hat{\mathbf{y}}={\rm Sigmoid}({\rm MLP}_r([\mathbf{r}_d''||\mathbf{r}_p'']))
         \end{equation}
         \noindent where $\mathbf{r}_d''$ and $\mathbf{r}_p''$ are the medical outputs of the prompted diagnosis and procedure sets, respectively. The parameters of ${\rm MLP}_r$ for hospital $h$ can be represented as $\Theta_{MLP}^{(h)}$.

         To maintain the general medical knowledge learned at the pretraining stage, we freeze the diagnosis and procedure embedding matrix and medical encoder when tuning the model. The prompt embedding is updated to learn the specific information for each hospital. The MLP is updated for precise recommendations. We tune and save the parameters of the prompt embedding and MLP on $\mathcal{R}^{(h)}$ to recommend for the hospital $h$. It is worth noting that we conduct inference for hospital $h$ using corresponding parameters of MLP $\Theta_{MLP}^{(h)}$, and prompt embedding $\mathbf{O}^{(h)}_d$ and $\mathbf{O}^{(h)}_p$.

    \subsection{Optimization and Inference}
    
        \textbf{\textit{Overall Optimization}}. 
         Based on Section \ref{sec_Pretraining} and \ref{sec_PromptTuning}, we summarize the optimization of TEMPT in Algorithm~\ref{alg_TEMPT}. Firstly, we pretrain the model on multi-center records $\mathcal{V}$ (line 2). During the pretraining, we calculate the losses of two self-supervised tasks (line 4-5), and get the pretrained parameters of embedding matrics and medical encoder (line 6). Then, based on the pretrained parameters, we conduct the prompt tuning (line 11) on each hospital. By calculating the loss of recommending medication (line 12), we get the hospital-specific parameters (line 13). We formulate the optimization problem as follows:

         \begin{align}
            \min_{\mathbf{E}_d,\mathbf{E}_p,\Theta_{encoder}}& \mathcal{L}_{pretrain} \tag{Pretraining Stage} \\
            \min_{\mathbf{o}_d, \mathbf{o}_p, \Theta_{MLP}}& \mathcal{L}_{tune} \tag{Prompt Tuning Stage}
        \end{align}

        \noindent \textbf{\textit{Inference}}. When inference, we recommend proper medications according to hospitals. Given a patient from the hospital $h$, we combine the pretrained parameter $\{\mathbf{E}_d, \mathbf{E}_p, \Theta_{encoder}\}$ and the hospital-specific parameters $\{\mathbf{O}^{(h)}_d, \mathbf{O}^{(h)}_p, \Theta_{MLP}^{(h)}\}$ to get the model. We follow the procedure of prompt tuning to give out the recommending probability as Equation~\eqref{eq12} shows. Then, we impose a threshold $t$ on the probability, \ie taking the medication whose probability is greater than $t$ into the recommended medication combination.

\section{Experiments} \label{sec_exp}

In this section, we will introduce the comprehensive experiments that are conducted on a real-world multi-center medical dataset. The analysis will be given out from the aspects according to the following Research Problems (\textbf{RQ}):

\begin{itemize}[leftmargin=*]
    \item \textbf{RQ1:} How does the proposed TEMPT perform compared with the state-of-the-art medication recommendation and multi-domain recommendation models? 
    \item \textbf{RQ2:} How does each component in TEMPT affect the recommending performance?
    \item \textbf{RQ3:} How do other parameter-efficient finetuning methods perform, compared with the proposed prompt tuning?
    \item \textbf{RQ4:} Does the proposed TEMPT have more storage and computation efficiency?
    \item \textbf{RQ5:} How do hyper-parameters affect the performance?
    \item \textbf{RQ6:} Can TEMPT serve better for small hospitals?
\end{itemize}

\begin{table}[]
\centering
\caption{The statistics of pre-processed eICU dataset.}

\resizebox{0.75\linewidth}{!}{%
\begin{tabular}{ll}
\hline
\textbf{Item} & \textbf{Number} \\ \hline
\# of records / hospitals                    &  102,363 / 80      \\
avg. / max / min \# of records per hospital  &  1,279.45 / 4,402 / 421     \\
diag. / prod. / med. vocabulary size         &  1,161 / 1,000 / 300      \\ 
avg. / max \# of diagnosis per records       &  3.44 / 74         \\
avg. / max \# of procedures per records      &  7.21 / 133     \\
avg. / max \# of medications per records     &  22.31 / 103     \\
\# of records in train / validation / test    &  81,884 / 10,235 / 10,237     \\
\hline
\end{tabular}
\label{eicu}}
\end{table}

\subsection{Experiment Settings}
\subsubsection{\textbf{Dataset}}
    We conduct experiments on eICU Collaborative Research Database\footnote{\url{https://eicu-crd.mit.edu/about/eicu/}}~\cite{pollard2018eicu}, which is a real-world multi-center medical dataset. It collects the data of patients admitted to critical care units in 2014 and 2015 from several hospitals. As far as we know, it is the \textbf{only} public dataset that can be used to research multi-center medication recommendation. We first remove the diagnosis, procedures and medications without valid codes. Then, for easier training and analysis~\cite{wu2022conditional}, we retain the top 300 medications and top 1,000 procedures according to the count of occurrence. To guarantee enough instances of each hospital for training and testing, we remove the hospital with under 400 records. Finally, we divide the data into train/validation/test by the ratio of 8:1:1 for each hospital. After pre-processing, the statistics of the dataset are listed in Table~\ref{eicu}. The eICU~\cite{pollard2018eicu} dataset was released in 2018, which collects the records of critical care patients from 208 hospitals. Compared with the most used dataset MIMIC-$\uppercase\expandafter{\romannumeral3}$~\cite{johnson2016mimic} in existing medication recommendation works~\cite{shang2019gamenet,yang2021safedrug,wu2022conditional}, eICU contains data from multi-center and can support our research. As for the number of records, MIMIC-$\uppercase\expandafter{\romannumeral3}$ has over 5,000 records for a single hospital, while most hospitals have under 1,500 records in eICU. It means that eICU has much more sparsity and is more difficult to well-train a medication recommendation model. In our work, we use the diagnosis.csv, treatment.csv, medication.csv and patient.csv in the eICU database to get the input diagnosis, procedures, and true prescribed medications.

\subsubsection{\textbf{Baseline Models}}

    We compare the performance of our model with the following three groups of baseline models: (i) medication recommendation models, (ii) LLM-based Models, (iii) multi-domain recommendation models, and (iv) some variant models of the proposed TEMPT. 

    \noindent\textit{\textbf{Medication Recommendation Models.}} Due to most models in this field devised for the multi-visit situation, we modify them as little as possible to fit our setting. It is worth noting that the input of all medication recommendation baselines is the same as ours, \ie diagnosis and procedure set, except for Leap~\cite{zhang2017leap}. The competing baselines include Leap~\cite{zhang2017leap}, GAMENet~\cite{shang2019gamenet}, G-Bert~\cite{shang2019pre} and COGNet~\cite{wu2022conditional}. 
    
    \begin{itemize}[leftmargin=*]
        \item Leap~\cite{zhang2017leap}. Leap models medication recommendation as a problem of sequential decision-making and adopts reinforcement learning to tune the parameters. We implement it as the original paper and only input the model with the diagnosis set. 
        
        \item GAMENet~\cite{shang2019gamenet}. It uses memory networks and graph neural networks to model the co-occurrence and interactions between medications. Since the medication in eICU is encoded by HICL code, the drug-drug interaction graph (DDI) in the original setting is not available. Therefore, we remove the DDI graph and dynamic memory module in GAMENet. 

        \item G-Bert~\cite{shang2019pre}. It pretrains the medication and diagnosis encoders on all of the single-visit records and then finetunes the model on multi-visit data. 
        However, under single-visit condition, no historical medication can be used as input. To make full use of the information, we substitute the procedures for medications as inputting with diagnosis. 

        \item COGNet~\cite{wu2022conditional}. COGNet considers recommending for the current visit as combining predicting new ones and copying from historical records. 
        {We remove the copy module in COGNet to adapt to the instance-based setting.}
        
    \end{itemize}

    \noindent\textit{\textbf{LLM-based Models}}. Large language models (LLMs) have revolutionized the research pattern of several fields, including recommender systems~\cite{wu2023survey}. To further verify the effectiveness of our TEMPT, we compare two LLM-based baselines.

    \begin{itemize}[leftmargin=*]
        \item GPT-4. Wornow~\etal~\cite{wornow2024zero} have explored that LLMs can conduct the zero-shot clinical trial matching task well given patients' EHR. Following this work, we organize the EHR into texts and adopt the powerful GPT-4 model to give out the recommended medications. The input prompt consists of the diagnosis and procedures of patients' current ICU visits, so the prompt template is as follows, ``\textit{In this ICU visit, the patient has diagnoses: <DIAG NAME>, ..., <DIAG NAME>; procedures: <PROC NAME>, ..., <PROC NAME>. Then, the patient should be prescribed: }''. The ``<DIAG NAME>'' and ``<PROC NAME>'' are the medical terms for diagnosis and procedure, respectively. The response from LLMs should be prescribed medication names.
        
        \item TALLRec~\cite{bao2023tallrec}. Instead of calling the API of closed-source LLMs, TALLRec supervised finetunes LLaMA-7B~\cite{touvron2023llama} to better adapt LLMs to fit for recommendation task. We follow the implementation of TALLRec and adopt the same prompt used for the GPT-4 baseline. 
    \end{itemize}

    \noindent\textit{\textbf{Multi-domain Recommendation Models.}} {To comprehensively position our proposed TEMPT in current studies,} we also compare several state-of-arts multi-domain recommendation models, including MMOE~\cite{ma2018modeling}, PLE~\cite{tang2020progressive} and STAR~\cite{sheng2021one}. The MMOE and PLE are multi-task based models and can be compatible with many neural architectures, so we apply the proposed medical encoder to them. In terms of the STAR baseline, we implement it according to the original paper completely. 
    For better performance, we train these models following the training scheme in \cite{sheng2021one}, \ie sampling mini-batch instances for each hospital.
    
    \begin{itemize}[leftmargin=*]
        \item MMOE~\cite{ma2018modeling}. MMOE designs shared expert networks for modeling various tasks and utilizes the gate networks to control the contribution of each expert. As for the implementation, we regard recommendations for hospitals as multiple tasks. However, since we have 80 hospitals, it is impractical to set up the same number of experts due to the time and space costs. Considering the output dimension of the expert network, we set 15 experts and the structure of each expert is the same as the medical encoder mentioned in Section~\ref{sec_model}.

        \item PLE~\cite{tang2020progressive}. Based on MMOE, PLE separates the experts into specific experts and shared experts to relieve the problem of detrimental interactions among different tasks. For the same reason as MMOE, we set up one shared expert and 15 specific experts, which means that each specific expert will correspond to 5 to 6 hospitals.

        \item STAR~\cite{sheng2021one}. It devises a centered network to learn the shared knowledge and domain-specific networks to capture the specific information of each domain. We implement it as illustrated in the original paper.
        
    \end{itemize}
    
    \noindent\textit{\textbf{Variant Models of TEMPT.}} To illustrate the efficiency between different training schemes, we build up three variant models equipped with the same model structure as the proposed TEMPT:
    \begin{itemize}[leftmargin=*]
        \item \textbf{Full-Train.} This model is equipped with the embedding table, medical encoder and the MLP to recommend medications. We train it on the whole multi-center dataset.
        
        \item \textbf{Single-Train.} The model has the same architecture as Full-Train. We train it for every hospital on their own records.

        \item \textbf{TEMPT (finetune)} This variant has the same pretraining stage as TEMPT, but it tunes all the parameters and is not equipped with the prompt vector.
    \end{itemize}
    
\subsubsection{\textbf{Implementation Details}} Our implementation is based on PyTorch 1.12 and Python 3.9.6. We train and test the codes on an AMD EPYC 7543 platform with RTX 3090 GPUs.  For G-Bert and the proposed TEMPT, the pretrained model and prompt-tuned model are selected by the masked diagnosis prediction and medication recommendation performance, respectively. The other baselines all select hyper-parameters based on the validation performance of medication recommendation.  We use Adam optimizer and the early-stopping strategy during the training. To ensure the reproducibility of the model and stability of the results, we repeatedly divide the dataset 5 times using the random seeds in $\{42,43,44,45,46\}$ according to Section 4.1 and run the model also with these seeds. The reported results in this paper are all averaged on five runs. For recommendation, we use the threshold $t=0.3$, because this value performs better in total.
However, due to the high costs of calling the API of GPT-4 and the extreme training consumption of finetuning LLaMA, we do not conduct repetitive experiments for the LLM-based baselines. 
The implementation code is available to ease reproducibility\footnote{\url{https://github.com/Applied-Machine-Learning-Lab/TEMPT}}. 

\subsubsection{\textbf{Evaluation Metrics}} We follow the previous works~\cite{shang2019gamenet,shang2019pre,yang2021safedrug,wu2022conditional} to use the Jaccard Similarity Score (denoted as Jaccard), Average F1 Score (denoted as F1) and Precision-Recall AUC (denoted as PRAUC) as the evaluation metrics. The reported metrics in the following are averaged over all of the records and hospitals. 

\begin{itemize}[leftmargin=*]
        \item \textbf{PRAUC}, which represents precision-recall area under curve. As mentioned in Section 3, we can get the recommending probability for each medication given a patient record. If the threshold is $t$, then the recommended medications for record $i$ are $\mathcal{\hat{M}}_i=\{m_j | \hat{y}_j >t\}$. Based on the recommended medication set, we get the definition of precision and recall: 
        \begin{equation} \label{eq14}
            {\rm Precision}_i=\frac{|\mathcal{M}_i \cap \hat{\mathcal{M}}_i|}{|\hat{\mathcal{M}}_i|}
        \end{equation}
        \begin{equation} \label{eq15}
            {\rm Recall}_i=\frac{|\mathcal{M}_i \cap \hat{\mathcal{M}}_i|}{|\mathcal{M}_i|}
        \end{equation}
        
        We sum the products of precision and recall under all possible thresholds and then get the PRAUC:
        \begin{equation} \label{eq16}
            {\rm PRAUC}=\sum_{k=1}^{|\mathcal{M}|} {\rm Precision}_i \cdot {\rm Recall}_i
        \end{equation}

        \item \textbf{Jaccard}, which evaluates the similarity between recommended and true medication sets. It is defined as follows:
        \begin{equation} \label{eq17}
            {\rm Jaccard}=\frac{|\mathcal{M}_i \cap \hat{\mathcal{M}}_i|}{|\mathcal{M}_i \cup \hat{\mathcal{M}}_i|}
        \end{equation}

        \item \textbf{F1}, which combines the precision and recall. It is defined as:
        \begin{equation}
            {\rm F1}=\frac{2 \cdot {\rm Precision}_i \cdot {\rm Recall}_i}{{\rm Precision}_i + {\rm Recall}_i}
        \end{equation}
    \end{itemize}

\begin{table}[!t]
\centering
\caption{Overall performance on eICU dataset. The boldface refers to the highest score and the underline indicates the best result of the baselines. ``\textbf{{\Large *}}'' indicates the statistically significant improvements (\ie two-sided t-test with $p<0.05$) over the best baseline. ``-'' means that the PRAUC is not applicable to LLM-based baselines, because they do not output the probability of each medication.}
\resizebox{0.9\linewidth}{!}{
\begin{tabular}{lccc}
\hline
                  \textbf{Models}& \textbf{PRAUC} & \textbf{Jaccard} & \textbf{F1-score} \\ \hline
Leap              & 0.3289 $\pm$ 0.0012 & 0.0739 $\pm$ 0.0003 & 0.1351 $\pm$ 0.0005    \\
GAMENet           & 0.4849 $\pm$ 0.0010 & 0.2778 $\pm$ 0.0015 & 0.4187 $\pm$ 0.0022    \\
G-Bert            & 0.4914 $\pm$ 0.0017 & 0.2913 $\pm$ 0.0015 & 0.4353 $\pm$ 0.0018    \\
COGNet            & 0.4890 $\pm$ 0.0012 & 0.2867 $\pm$ 0.0013 & 0.4279 $\pm$ 0.0015    \\ \hline
GPT-4 & - & 0.0669 & 0.1246 \\
TALLRec & - & 0.2037 & 0.3207 \\
\hline
MMOE              & 0.5188 $\pm$ 0.0013 & \underline{0.3075 $\pm$ 0.0021} & \underline{0.4510 $\pm$ 0.0023}    \\
PLE               & 0.5060 $\pm$ 0.0017 & 0.3033 $\pm$ 0.0008 & 0.4480 $\pm$ 0.0011    \\
STAR              & \underline{0.5212 $\pm$ 0.0005} & 0.3068 $\pm$ 0.0011 & 0.4509 $\pm$ 0.0013    \\ \hline
Full-Train        & 0.4840 $\pm$ 0.0014 & 0.2839 $\pm$ 0.0010 & 0.4265 $\pm$ 0.0011     \\
Single-Train      & 0.5033 $\pm$ 0.0009 & 0.2950 $\pm$ 0.0018 & 0.4364 $\pm$ 0.0011   \\ \hline
\textbf{TEMPT (finetune)} & 0.5407 $\pm$ 0.0011 & 0.3300 $\pm$ 0.0012 & 0.4778 $\pm$ 0.0015 \\
\textbf{TEMPT}              & \textbf{0.5468 $\pm$ 0.0016{\Large *}} & \textbf{0.3318 $\pm$ 0.0010{\Large *}} & \textbf{0.4790 $\pm$ 0.0013{\Large *}}    \\ \hline
\end{tabular}
\label{exp-overall}}
\end{table}

\subsection{Overall Performance Comparison (RQ1)}

    As Table~\ref{exp-overall} shows, we repeatedly conduct five experiments with different random seeds on each model and report the averaged performance and standard deviation. Overall, the proposed TEMPT outperforms all competing baselines on three metrics. Besides, we can get several more detailed conclusions from the results as follows: 
    \begin{itemize}[leftmargin=*]
        \item Observing the performance of compared medication recommendation models, they are much worse than the MDR models and the proposed TEMPT. This is because the existing medication recommendation models ignore the situation of multi-center. The problem of the significant difference between hospitals results in the difficulty of training a unified model for all hospitals.

        \item Among the medication recommendation models, Leap performs inferior to others, because it only uses the information of diagnosis. It is also worth noting that G-Bert is a pretrain-based model and outperforms others a little, which illustrates that inner medical knowledge can benefit recommending medications for various hospitals.

        \item From the results, we can find that both GPT-4 and TALLRec underperform the proposed TEMPT. The results align with an existing research paper~\cite{dai2023uncovering}, which indicates that LLMs are better at the zero-shot or few-shot setting while inferior to traditional recommendation models with sufficient training data. Furthermore, the general LLMs, \eg GPT-4 and LLaMA, do not have enough medical knowledge, leading to unsatisfied performance for medication recommendation, a typical medical application.

        \item The experimental results also show that MDR models can get the second-best performance. We believe that it is because MDR can learn various distributions of hospitals, which shows the importance of researching the multi-center situation. However, MMOE and PLE still cannot overrun the proposed TEMPT, because they are not fit for too many domains. In general, they need the same number of expert networks with domains, but it is impossible when we have 80 hospitals. Besides, the PLE even has to share a specific expert with some of the hospitals due to too many hospitals, so the performance degrades much.

        \item As for STAR, though it can be applied to conditions with many domains, it cannot be compatible with more advanced neural architectures, which limits its performance.

        \item Single-Train and Full-Train only learn the specific knowledge of each hospital and the general knowledge of all hospitals, respectively. The results show that both of them cannot satisfy the multi-center situation.

        \item A variation of the TEMPT uses common finetuning rather than the proposed prompt, denoted as TEMPT (finetune). It outperforms other medication recommendation and MDR models, which illustrates two-fold facts: (i) pretrain-finetune is a more efficient scheme for multi-center medication recommendation. (ii) the proposed two self-supervised tasks help the model capture useful general medical information. Furthermore, the proposed TEMPT is better than the variation, because it can obtain more specific information about each hospital and relieve the catastrophic forgetting issue via prompt tuning.
        
    \end{itemize}
    
    \noindent In summary, we can conclude that the proposed TEMPT outperforms the competing state-of-the-art medication recommendation and MDR models as the answer to \textbf{RQ1}.

\subsection{Ablation Study (RQ2)}

\begin{table}[!t]
\centering
\caption{Ablation study for different components of TEMPT on eICU. The boldface refers to the highest score and the underline indicates the best result of the competitors. ``\textbf{{\Large *}}'' indicates the statistically significant improvements (\ie two-sided t-test with $p<0.05$) over the competitor.}
\resizebox{0.85\textwidth}{!}{
\begin{tabular}{lccc}
\hline
                 \textbf{Models}& \textbf{PRAUC} & \textbf{Jaccard} & \textbf{F1-score} \\ \hline
TEMPT-DE   & 0.5424 $\pm$ 0.0014 & 0.3267 $\pm$ 0.0020 & 0.4730 $\pm$ 0.0021   \\
TEMPT-SM   & 0.5437 $\pm$ 0.0014 & 0.3269 $\pm$ 0.0020 & 0.4732 $\pm$ 0.0026   \\  \hline
TEMPT (finetune) w/o MP   & 0.5032 $\pm$ 0.0079 & 0.2947 $\pm$ 0.0100 & 0.4354 $\pm$ 0.0113   \\
TEMPT (finetune) w/o CL   & 0.5383 $\pm$ 0.0007 & 0.3281 $\pm$ 0.0005 & 0.4753 $\pm$ 0.0003   \\ 
TEMPT (finetune)  & 0.5407 $\pm$ 0.0011 & \underline{0.3300 $\pm$ 0.0012} & \underline{0.4778 $\pm$ 0.0015}   \\ \hline
TEMPT w/o MP     & 0.4809 $\pm$ 0.0089 & 0.2719 $\pm$ 0.0105 & 0.4116 $\pm$ 0.0078   \\
TEMPT w/o CL      & \underline{0.5445 $\pm$ 0.0013} & 0.3275 $\pm$ 0.0012 & 0.4738 $\pm$ 0.0014   \\ \hline
TEMPT             & \textbf{0.5468 $\pm$ 0.0016}* & \textbf{0.3318 $\pm$ 0.0010}* & \textbf{0.4790 $\pm$ 0.0013}    \\ \hline
\end{tabular}
}
\label{tab_exp_ablation}
\end{table}

    To respond to the \textbf{RQ2}, we conduct ablation experiments on eICU and the results are shown in Table~\ref{tab_exp_ablation}. 
    \begin{itemize}[leftmargin=*]
        \item At first, we analyze the effectiveness of the shared medical encoder and separate MLP. \textit{TEMPT-DE} represents we use separate medical encoders for diagnosis and procedure, respectively. \textit{TEMPT-SM} is that we use the same MLPs for the pre-training tasks. From the results, we can conclude that using the shared medical encoder and separate MLPs can get the best performance. This shows that it is better to consider diagnosis and procedure as two independent tasks and use the shared medical encoder to learn similar medical knowledge.
        \item Then, we discuss how the pretrain tasks affect the performance of TEMPT. \textit{TEMPT (finetune) w/o MP} represents that we finetune the model only pretrained by contrastive task, while \textit{TEMPT (finetune) w/o CL} for the model only pretrained by mask prediction task. \textit{TEMPT w/o MP} and \textit{TEMPT w/o CL} represent the proposed model without mask prediction and contrastive self-supervised task. The experimental results show that removing any of the pretrain tasks will cause a performance decrease in both finetuning and prompt tuning, which illustrates that the two pretrain tasks can help the model learn general medical knowledge. More performance decreases on the condition without mask prediction, because of the problem of false negatives~\cite{saunshi2019theoretical} in our view. We will explore it in future work.
        \item Also, the recommending performance of TEMPT falls when we use the finetuning method rather than the proposed hospital-based prompt tuning. This phenomenon illustrates that prompt tuning relieves the problem of catastrophic forgetting and captures the specific information of each hospital's distribution.
    \end{itemize}

\subsection{Analysis for Parameter-efficient Finetuning (RQ3)}

There are several parameter-efficient finetuning (PEFT) methods that can be adapted to our TEMPT. To verify the effectiveness of the proposed prompt tuning (RQ3), we compare four types of popular PEFT methods and show their performance in Table~\ref{tab_peft}.

\begin{itemize}[leftmargin=*]
    \item \textit{TEMPT (Adapter)} signifies that we employ the same pretraining process as TEMPT, but use the adapter~\cite{houlsby2019parameter} for finetuning. The adapter is integrated into each transformer-based medical encoder of our TEMPT model. The performance comparison reveals that prompt tuning can outperform the adapter. This could be attributed to the fact that the adapter often necessitates more training data~\cite{abed2024multi}, but in our multi-center settings, most hospitals have no more than 2,000 samples.
    
    \item Regarding \textit{TEMPT (LoRA)}, we implement it using the same pretraining process, but finetune the model by pairing each linear layer in the transformer-based medical encoder with a set of low-rank metrics~\cite{hu2021lora}. The results in Table~\ref{tab_peft} suggest that LoRA is less effective than our proposed prompt tuning method. LoRA is a type of reparameterization method~\cite{lialin2023scaling}, which implies that it modifies the parameters of the linear layers in the original model with a smaller update step. However, updating too many parameters of the original model may lead to the issue of catastrophic forgetting~\cite{ramasesh2022effect}, resulting in suboptimal performance, akin to full finetuning.
    
    \item Prefix tuning introduces prompt vectors into each transformer layer, whereas prompt tuning only inserts them into the input layer. In the experiments, \textit{TEMPT (Prefix)} denotes that we employ the same pretraining process as TEMPT, but use prefix tuning for finetuning. Our results indicate that our prompt method outperforms \textit{TEMPT (Prefix)}. This could be because \textit{TEMPT (Prefix)} has too many prompt parameters to learn, leading to underfitting. This observation aligns with our findings that the optimal prompt number is 2 in hyper-parameter experiments. 
    
    \item We also compare with P-Tuning~\cite{liu2023gpt}. \textit{TEMPT (P-Tuning-1)} inputs the pseudo prompts to the prompt encoder, and \textit{TEMPT (P-Tuning-2)} takes the hospital ID as the input of the prompt encoder. From the results, we find that both these variants achieve inferior performance than TEMPT. It is caused by the difficulty of learning a good prompt encoder, while direct prompt insertion of TEMPT is more effective in learning hospital-specific information. Besides, these three prompt tuning variants surpass \textit{TEMPT (finetune)}, indicating the effectiveness of the application of prompt tuning for multi-center medication recommendation.
\end{itemize}

\begin{table}[!t]
\centering
\caption{Experiments for comparing different parameter-efficient finetuning methods. The boldface refers to the highest score and the underline indicates the best result of the competitors. ``\textbf{{\Large *}}'' indicates the statistically significant improvements (\ie two-sided t-test with $p<0.05$) over the competitor.}
\resizebox{0.8\textwidth}{!}{
\begin{tabular}{lccc}
\hline
                 \textbf{Models}& \textbf{PRAUC} & \textbf{Jaccard} & \textbf{F1-score} \\ \hline
TEMPT (finetune)          & 0.5407 $\pm$ 0.0011 & \underline{0.3300 $\pm$ 0.0012} & \underline{0.4778 $\pm$ 0.0015}    \\ 
\midrule
TEMPT (Adapter)   & 0.5412 $\pm$ 0.0011 & 0.3266 $\pm$ 0.0010 & 0.4753 $\pm$ 0.0014   \\ 
TEMPT (LoRA)   & 0.5425 $\pm$ 0.0012 & 0.3270 $\pm$ 0.0009 & 0.4760 $\pm$ 0.0013   \\ 
TEMPT (Prefix)   & 0.5436 $\pm$ 0.0013 & 0.3292 $\pm$ 0.0008 & 0.4768 $\pm$ 0.0011   \\
TEMPT (P-Tuning-1)   & \underline{0.5438 $\pm$ 0.0012} & 0.3288 $\pm$ 0.0015 & \underline{0.4778 $\pm$ 0.0018}   \\ 
TEMPT (P-Tuning-2)   & 0.5434 $\pm$ 0.0012 & 0.3287 $\pm$ 0.0011 & 0.4761 $\pm$ 0.0013   \\ \hline
TEMPT             & \textbf{0.5468 $\pm$ 0.0016}* & \textbf{0.3318 $\pm$ 0.0010}* & \textbf{0.4790 $\pm$ 0.0013}    \\ 
\hline
\end{tabular}
}
\label{tab_peft}
\end{table}

\begin{figure}[!t]
\centering
    \begin{minipage}{0.49\linewidth}
		\centering
        \begin{subfigure}{1\linewidth}
		\includegraphics[width=0.7\linewidth]{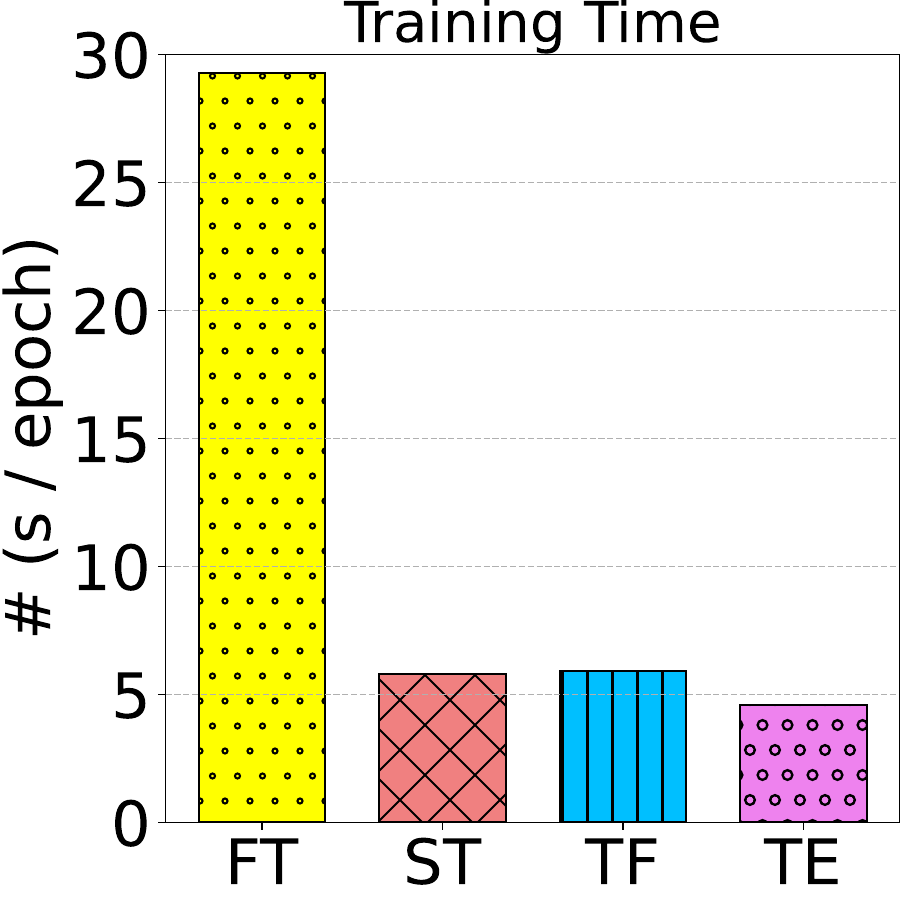}
		\caption{Computation cost}
		\label{fig_exp_eff_mem}
        \end{subfigure}
	\end{minipage}
    \begin{minipage}{0.49\linewidth}
		\centering
        \begin{subfigure}{1\linewidth}
		\includegraphics[width=0.7\linewidth]{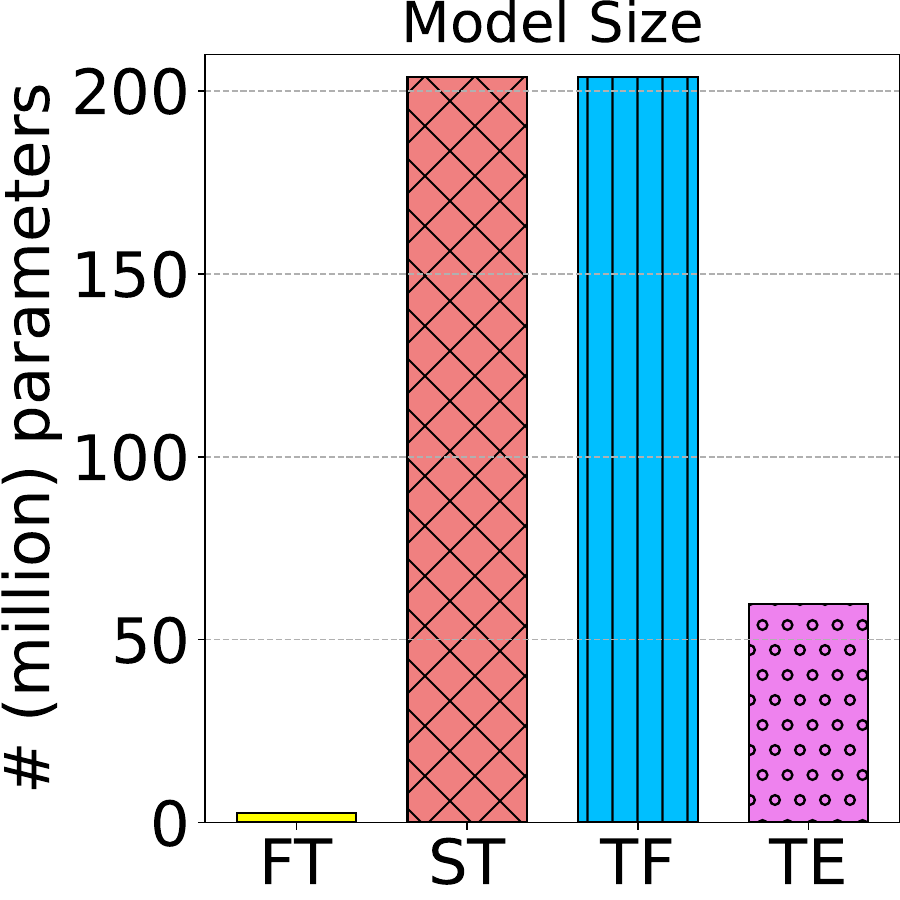}
		\caption{Storage cost.}
		\label{fig_exp_eff_sto}
        \end{subfigure}
	\end{minipage}
\caption{The computation and storage cost of each model.}
\label{fig_exp_eff}

\end{figure}

\subsection{Efficiency Analysis (RQ4)}

    In this subsection, we aim to explore the computation and storage efficiency of the proposed TEMPT. As shown in Figure~\ref{fig_exp_eff}, we compare the training time and model size between Full-Train (FT), Single-Train (ST), TEMPT (finetune) (TF) and TEMPT (TE). Full-Train consumes much more training time per epoch than the other three models, because it is trained on the whole multi-center data. Single-Train, TEMPT (finetune) and TEMPT are all trained on the records in a single hospital, but TEMPT is faster than the others. This is because TEMPT only needs to tune the prompt embedding and MLP in the model, while the other two models have to tune the whole model. As for the model size, Full-Train is the smallest because it serves all hospitals with one model, but it also causes poor performance. The model size of TEMPT is much smaller than the others, because all hospitals share one medical encoder and embedding layers, while the other two models have specific parameters of the whole for each hospital. To sum up, the proposed TEMPT has relatively higher computation and storage efficiency.

\begin{figure}[!t]
\begin{minipage}[t]{0.33\linewidth}
\centering
\begin{subfigure}{1\linewidth}
    \includegraphics[scale=0.25]{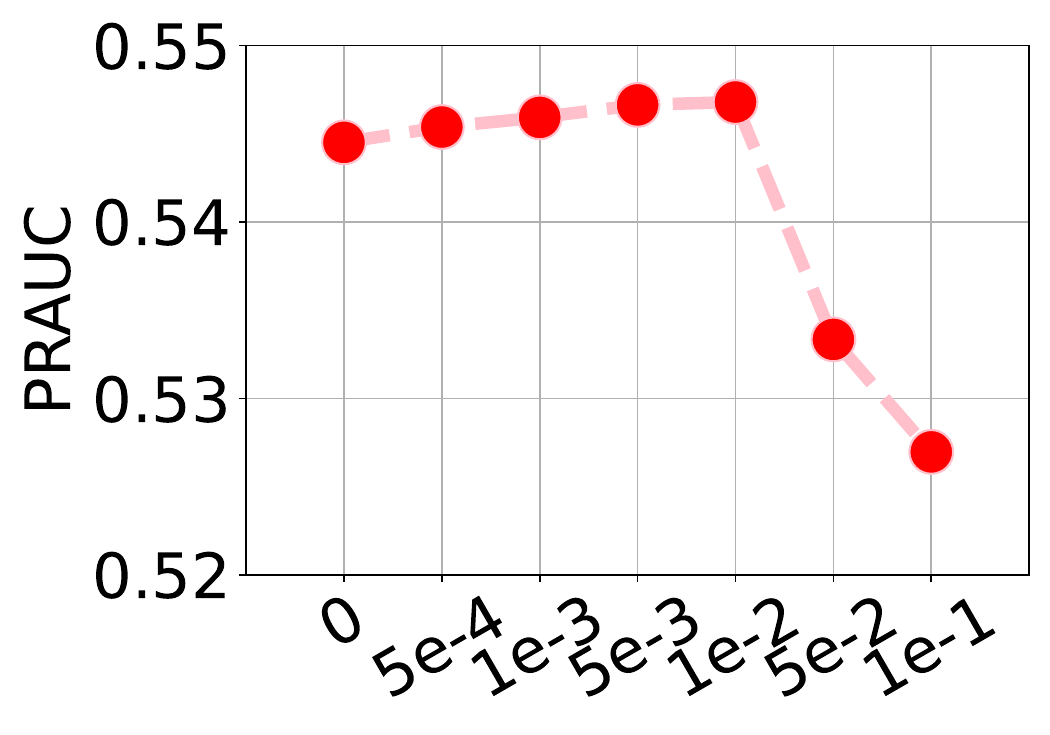}
    \caption{Weight of contrastive loss $\gamma$}
\label{fig_exp_hyper_gamma}
\end{subfigure}
\end{minipage}%
\begin{minipage}[t]{0.33\linewidth}
    \centering
\begin{subfigure}{1\linewidth}
    \includegraphics[scale=0.25]{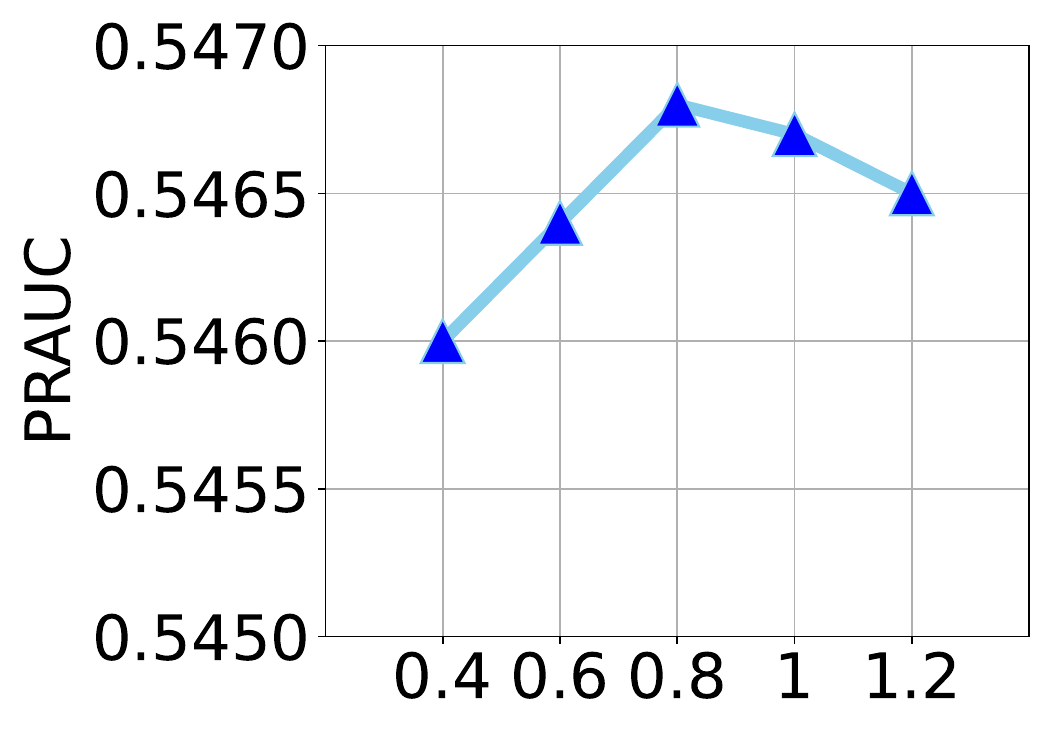}
    \caption{Temperature parameter $\tau$}
    \label{fig_exp_hyper_tau}
\end{subfigure}
\end{minipage}%
\begin{minipage}[t]{0.33\linewidth}
\centering
\begin{subfigure}{1\linewidth}
    \includegraphics[scale=0.25]{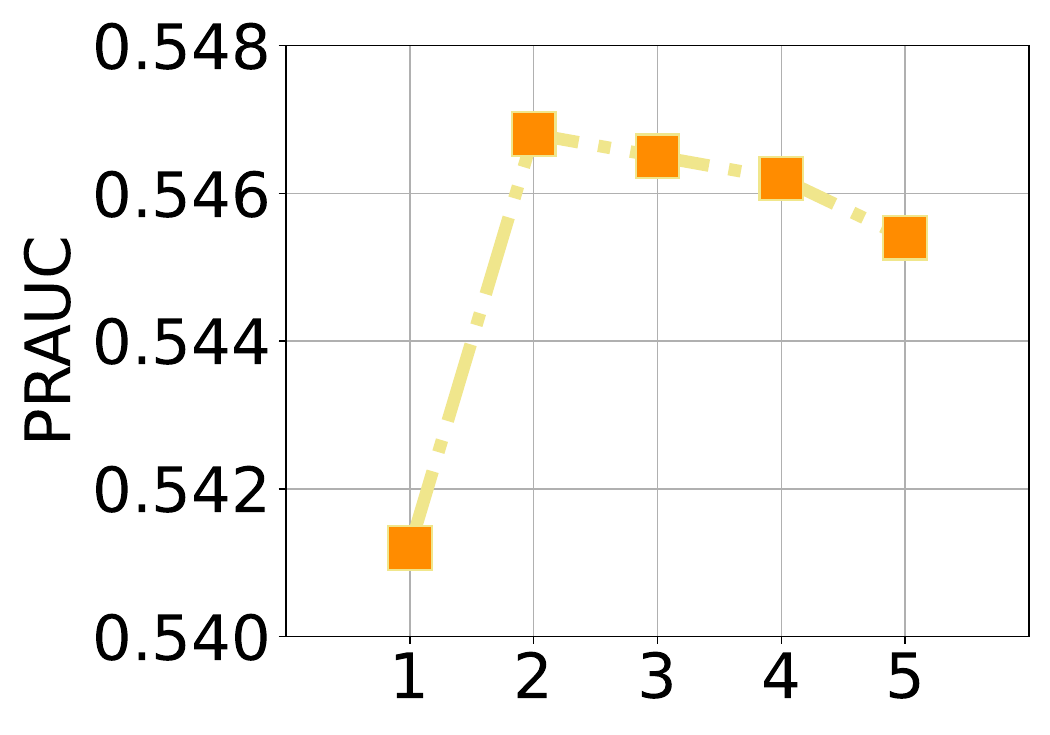}
    \caption{Number of prompt $b$}
\label{fig_exp_hyper_b}
\end{subfigure}
\end{minipage}
\caption{The performance of the model with various values of three hyper-parameters, \ie the weight of contrastive loss $\gamma$, the temperature parameter $\tau$ and the number of prompt $b$.}
\label{fig_exp_hyper}
\end{figure}

\subsection{Hyper-parameter Analysis (RQ5)} \label{sec_case}

    The proposed TEMPT owns three important hyper-parameters, \ie the weight of contrastive loss $\gamma$, the temperature parameter $\tau$ and the number of prompt $b$. To explore how they affect the performance of the model, we conduct experiments on various values of them. As shown in Figure~\ref{fig_exp_hyper}, the performance increases consistently as $\gamma$ varies from 0 to 1e-2, which illustrates that the contrastive task helps the model learn the inter-relationship between two inputs and thus improve the quality of representations. However, due to the larger magnitude of the contrastive loss, a continuous increase of $\gamma$ degrades the performance. As for the temperature parameter, the recommending performance first rises then falls when $\tau$ is ranged from $0.4$ to $1.2$, because the smaller temperature can promote the uniformity of representation distribution, but the too small value will cause the difficulty of optimization~\cite{wang2021understanding}. Besides, we seek the most proper number of prompts. The result shows that two prompts benefit the proposed model the most. When $b=1$, the performance is slightly better than without a prompt, because only one prompt vector is not enough to learn the specific information of each hospital. By comparison, more than two prompts are abundant for the model and cause a decrease in performance. To ease the reproduction of all the experimental results in this paper, we also show all the hyper-parameters used in our experiments in Table~\ref{tab_hyper}.

\begin{table*}[!t]
\centering
\caption{The hyper-parameter settings in our experiments.}
\resizebox{1\textwidth}{!}{
\begin{tabular}{lcccccccc}
\hline
\textbf{Hyper-parameter} & \textbf{Leap} &\textbf{GAMENet} & \textbf{G-Bert} & \textbf{COGNet} & \textbf{MMOE} & \textbf{PLE} & \textbf{STAR} & \textbf{TEMPT} \\ \hline
Embedding dimension $c$  & 300 & 300 & 300 & 300 & 300 & 300 & 300 & 300      \\
Batch size  & 64 & 64 & 64 & 64 & 64 & 64 & 64 & 64     \\  
Learning rate & 5e-4 & 5e-4 & - & 5e-4 & 5e-4 & 5e-4 & 5e-4 & - \\
Learning rate of pre-training  & - & - & 5e-4 & - & - & - & - & 5e-4    \\
Learning rate of tuning  & - & - & 5e-4 & - & - & - & - & 5e-4    \\ 
Transformer layers $K$ & - & - & 2 & 2 & 2 & 2 & -  &   2     \\ 
Number of multi-head attention & - & - & 2 & 2 & 2 & 2 & - &  2        \\
Max length of transformer inputs & - & - & 100 & 100 & 100 & 100 & - &  100     \\
Number of experts & - & - & - & - & 15 & 16 & - & - \\
Dimension of expert & - & - & - & - & 20 & 20 & - & - \\
Temperature parameter $\tau$  & - & - & - & - & - & - & - & 0.8  \\
Weight of contrastive loss $\gamma$  & - & - & - & - & - & - & - & 1e-2   \\
Number of prompt $b$   & - & - & - & - & - & - & - & 2   \\
\hline
\end{tabular}}
\label{tab_hyper}
\vspace{-3mm}
\end{table*}

\begin{table}[!t]
\centering
\caption{Performance comparison on different scales of hospitals. We divide the hospitals into three groups according to the amount of records, \ie small, medium and large hospitals. ``Impr'' indicates the ratio of performance improvement of TEMPT over Single-Train.}
\resizebox{0.95\textwidth}{!}{
\begin{tabular}{lcccc}
\hline
                  \textbf{Models}& \textbf{PRAUC}$_{small}$ & \textbf{PRAUC}$_{medium}$ & \textbf{PRAUC}$_{large}$ & \textbf{PRAUC}$_{all}$ \\ \hline
Leap    & 0.3293 & 0.3293 & 0.3272 & 0.3289 \\
GAMENet & 0.4797 & 0.4897 & 0.4901 & 0.4849 \\
G-Bert  & 0.4832 & 0.4998 & 0.4979 & 0.4914 \\
COGNet  & 0.4814 & 0.4941 & 0.4953 & 0.4890 \\ \hline
MMOE    & 0.5249 & 0.5147 & 0.5095 & 0.5188 \\
PLE     & 0.5115 & 0.5026 & 0.4971 & 0.5060 \\
STAR    & 0.5239 & 0.5186 & 0.5187 & 0.5212 \\ \hline
Full-Train        & 0.4766 & 0.4908 & 0.4913 & 0.4840   \\
Single-Train      & 0.5045 & 0.4996 & 0.5080 & 0.5033  \\ \hline
TEMPT (finetune)          & 0.5436 & 0.5380 & 0.5380 & 0.5407   \\
TEMPT (finetune-freeze)   & 0.5441 & 0.5375 & 0.5356 & 0.5404  \\ \hline
\textbf{TEMPT}     & \textbf{0.5509} & \textbf{0.5439} & \textbf{0.5406}& \textbf{0.5468}    \\ \hline
\textbf{Impr} & 9.20\% & 8.87\% & 6.42\% & 8.64\% \\
\hline
\end{tabular}
}
\label{exp-case}
\end{table}

\subsection{Analysis for Hospitals of Different Scales (RQ6)}

    One important reason to research multi-center medication recommendation is to assist small hospitals, which is much more meaningful for the healthcare system. As we know, small hospitals often own a small number of medical records, which easily causes the problem of underfitting.  Therefore, we experiment with the proposed models and competing baselines on different sizes of hospitals. We divide the hospitals into three groups: \textit{small} ($r \leq 1000$), \textit{medium} ($1000 < r \leq 2000$) and \textit{large} ($r > 2000$), where $r$ is the record count of each hospital.
    It is worth noting that the train-test split metric is the same as before. For evaluation, we test the models on each single hospital and then average the metric values based on the hospital groups.
    
    As Table~\ref{exp-case} shows, TEMPT outperforms other baselines overall and all three separated groups, which illustrates the efficiency of our TEMPT. In detail, we carefully get four conclusions as follows as the response to the \textbf{RQ6}:

    \begin{itemize}[leftmargin=*]
        \item We find that Full-Train not only performs worse overall, but also its accuracy in small hospitals is much lower. This phenomenon illustrates that the unified model tends to fit the most records from the large hospital and the small hospital records are ignored during training.

        \item Compared with Full-Train, Single-Train and most medication recommendation models can elevate performance in small hospitals. Single-Train specifies one model for each hospital and thus can benefit small hospitals. Other medication recommendation models elevate the performance by more fabricated architectural designs.

        \item MDR baselines even get better performance in small hospitals than large hospitals, which is different from Full-Train and Single-Train. It shows that capturing correlations between various hospitals can help alleviate the data scarcity problem for small hospitals.

        \item TEMPT and its variations outperform the other baselines on all hospital groups, because they have learned general medical knowledge that can enhance the model. It is worth noting that freezing parameters when finetuning is better than no freezing in small hospitals, which shows that the problem of catastrophic forgetting leads to a decrease in performance. In contrast, large hospitals need to fit their specific distributions more, so the performance is worse when freezing. TEMPT alleviates the problem of catastrophic forgetting and various distributions simultaneously, so it can get an increase in all three groups of hospitals.

        \item It is worth noting that, MDR baselines and TEMPT achieve more accuracy promotion on small hospitals, compared with the other models that do not consider multi-center. Typically, we can find that TEMPT surpasses Single-Train $9.20\%$ in the small hospital group, while the improvement is only $6.42\%$ for large hospitals. On the one hand, such a phenomenon shows that small hospitals suffer from the problem of little data. On the other hand, it can demonstrate the necessity to formulate the problem as a multi-center setting to benefit small hospitals, which is important for the whole healthcare system.
        
    \end{itemize}

    \noindent By analysis, the proposed TEMPT serves better for small hospitals than existing methods, due to its better combination of homogeneous and heterogeneous information between all hospitals.

\begin{figure}[htbp]
    \centering
    \begin{minipage}[b]{0.3\linewidth}
        \centering
        \begin{subfigure}{1\linewidth}
        \includegraphics[scale=0.27]{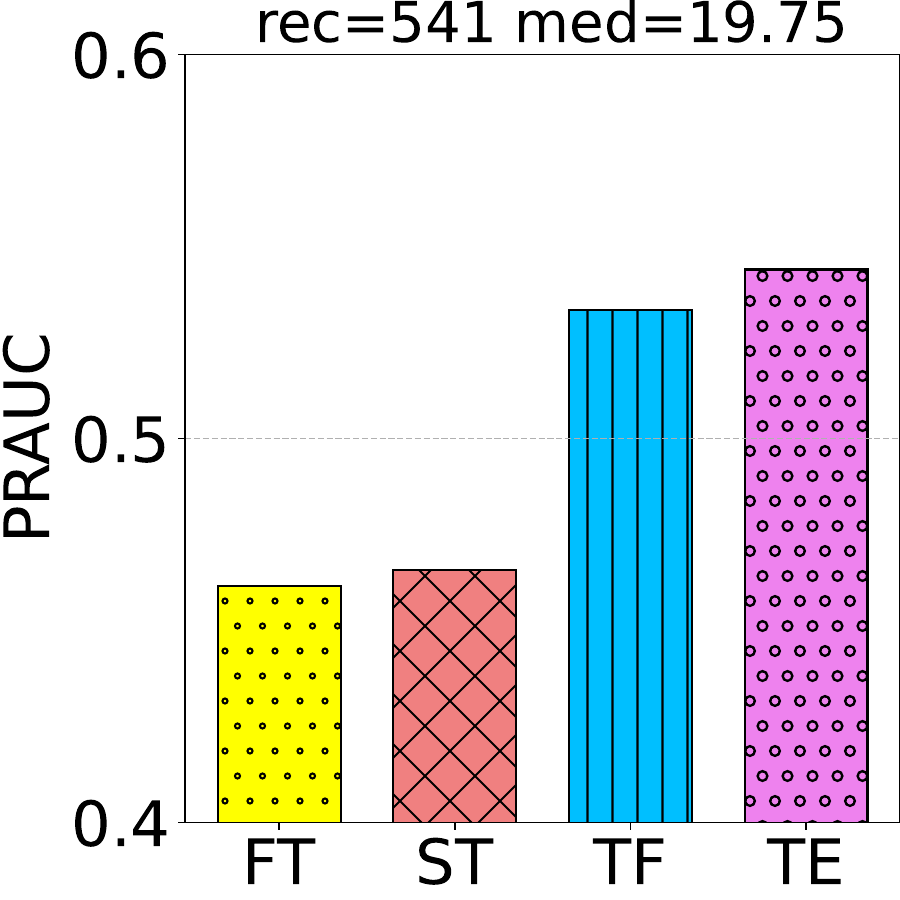}
        \caption{id=92}
    \end{subfigure}
    \end{minipage}%
    \begin{minipage}[b]{0.3\linewidth}
    \centering
    \begin{subfigure}{1\linewidth}
        \includegraphics[scale=0.27]{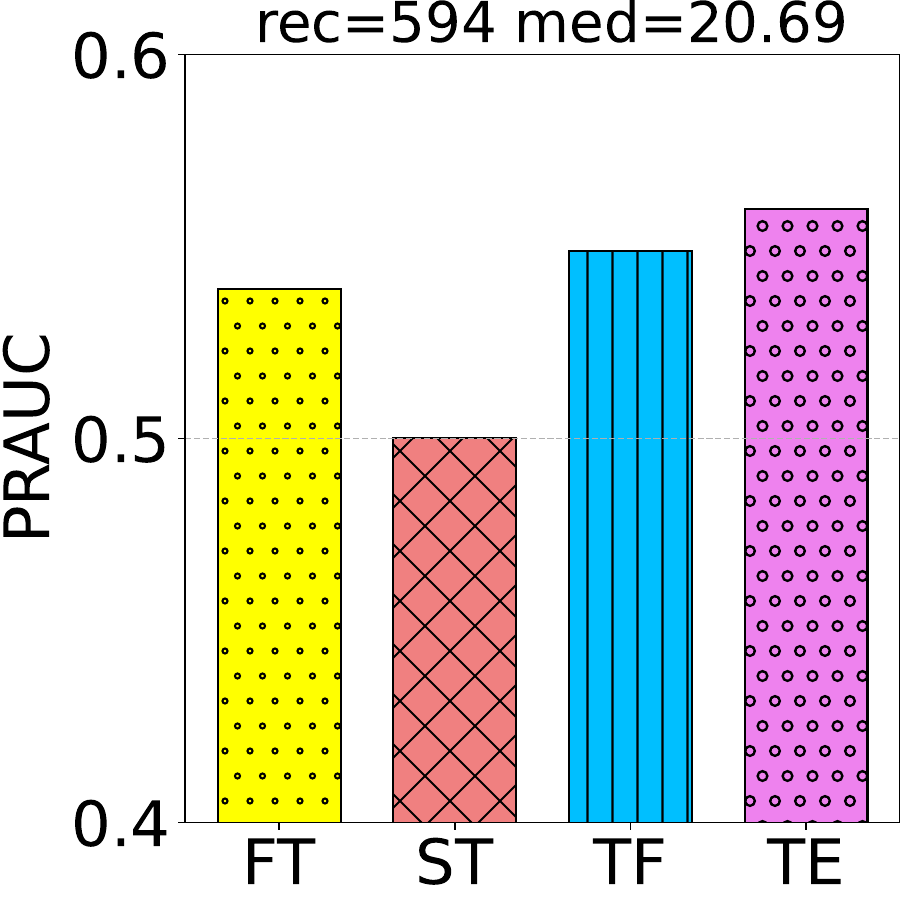}
        \caption{id=146}
    \end{subfigure}
    \end{minipage}%
    \begin{minipage}[b]{0.3\linewidth}
    \centering
    \begin{subfigure}{1\linewidth}
        \includegraphics[scale=0.27]{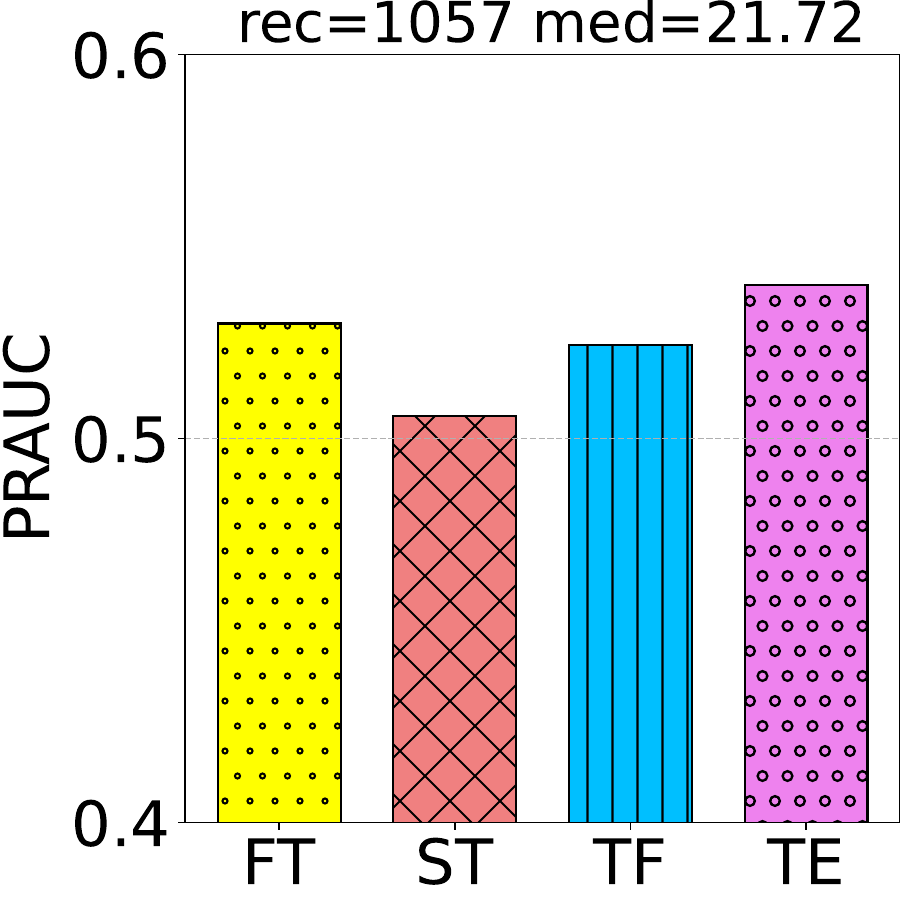}
        \caption{id=157}
    \end{subfigure}
    \end{minipage}%
    \\
    \begin{minipage}[b]{0.3\linewidth}
    \centering
    \begin{subfigure}{0.95\linewidth}
        \includegraphics[scale=0.27]{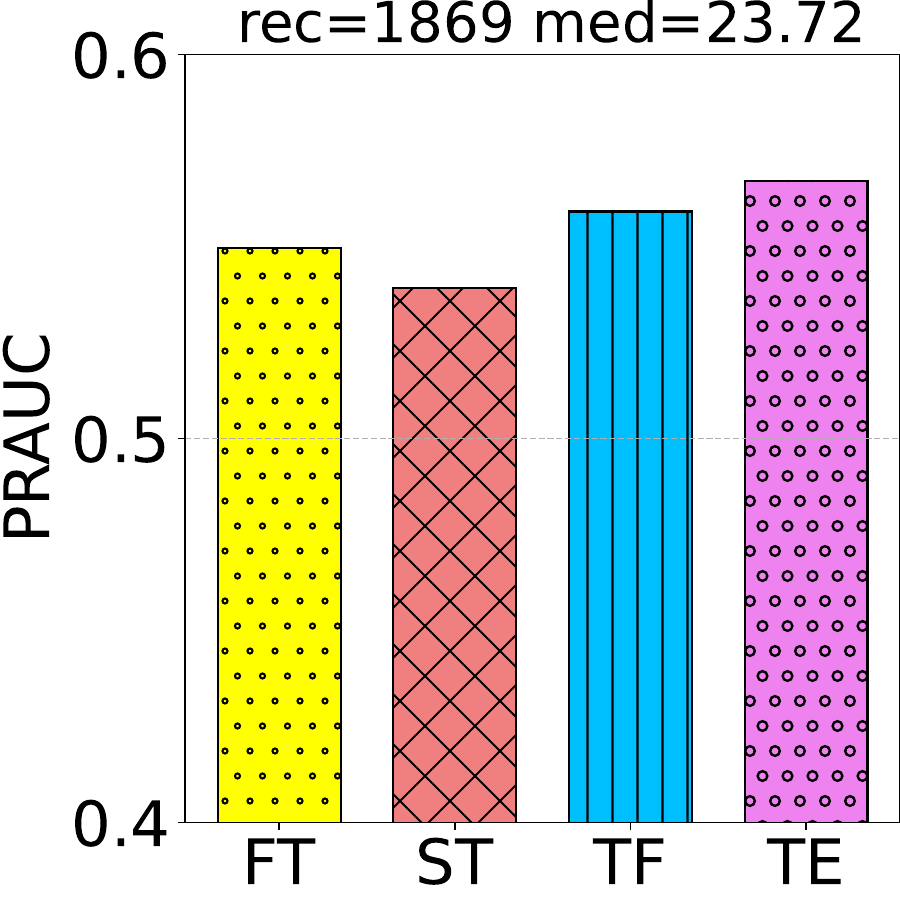}
        \caption{id=176}
    \end{subfigure}
    \end{minipage}
    \begin{minipage}[b]{0.3\linewidth}
        \centering
        \begin{subfigure}{0.95\linewidth}
        \includegraphics[scale=0.27]{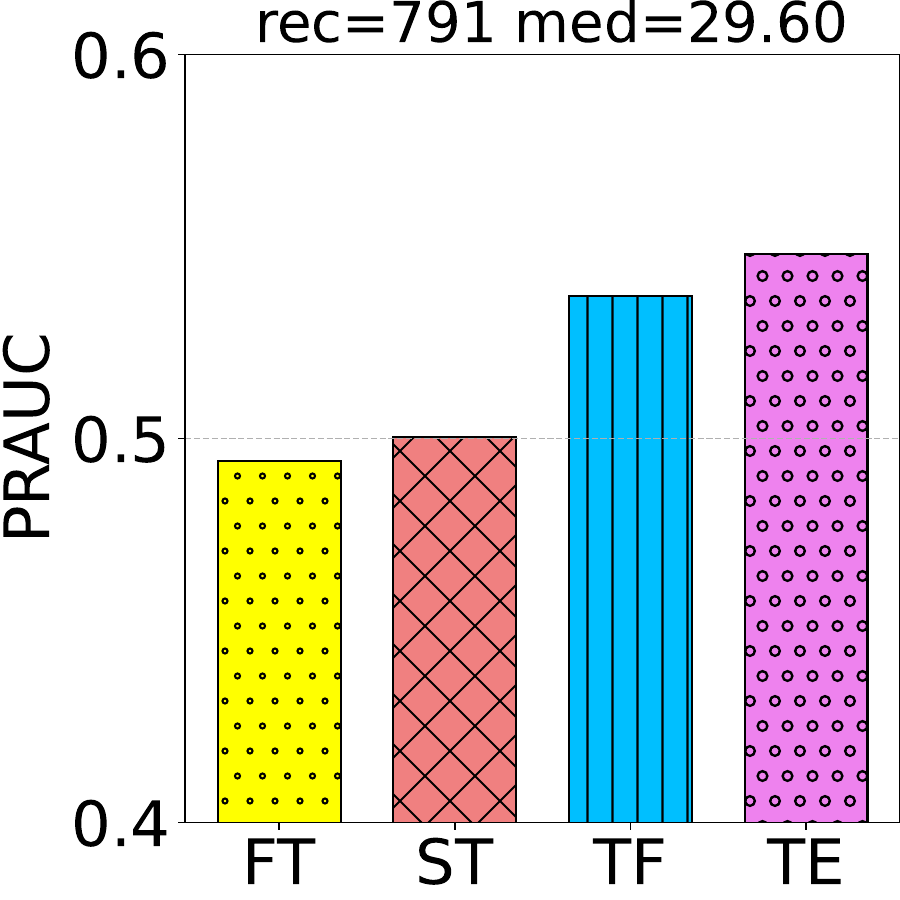}
        \caption{id=181}
    \end{subfigure}
    \end{minipage}%
    \begin{minipage}[b]{0.3\linewidth}
    \centering
    \begin{subfigure}{0.95\linewidth}
        \includegraphics[scale=0.27]{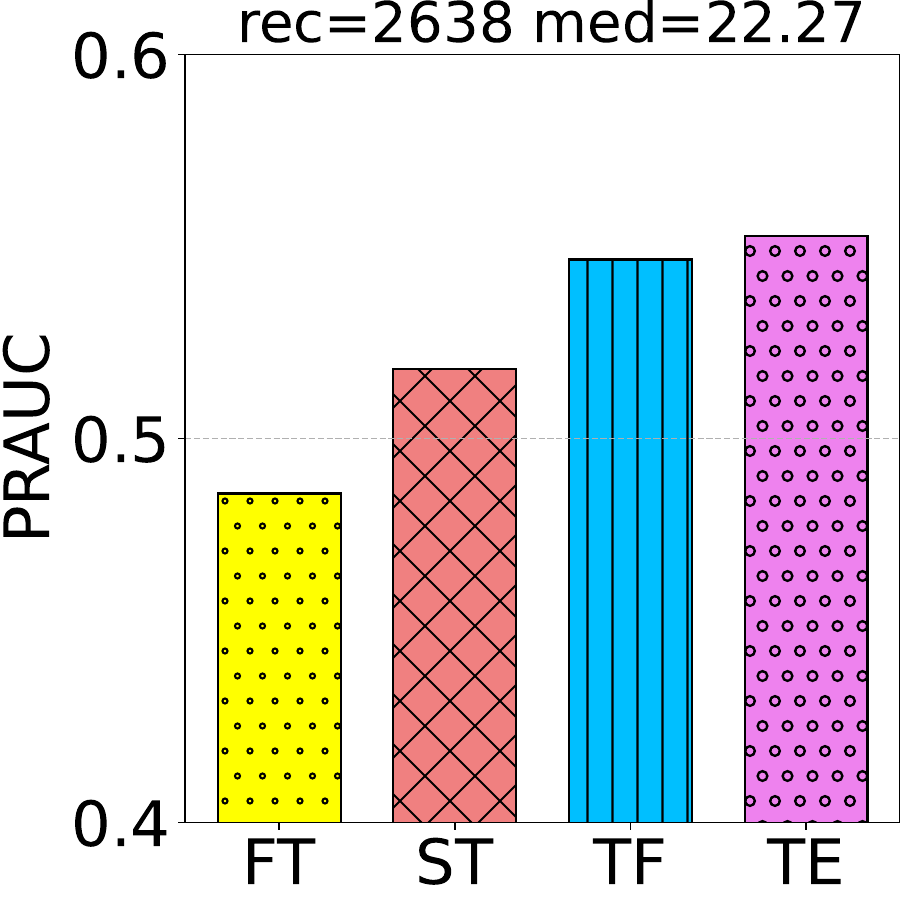}
        \caption{id=199}
    \end{subfigure}
    \end{minipage}%
    \\
    \begin{minipage}[b]{0.3\linewidth}
    \centering
    \begin{subfigure}{0.95\linewidth}
        \includegraphics[scale=0.27]{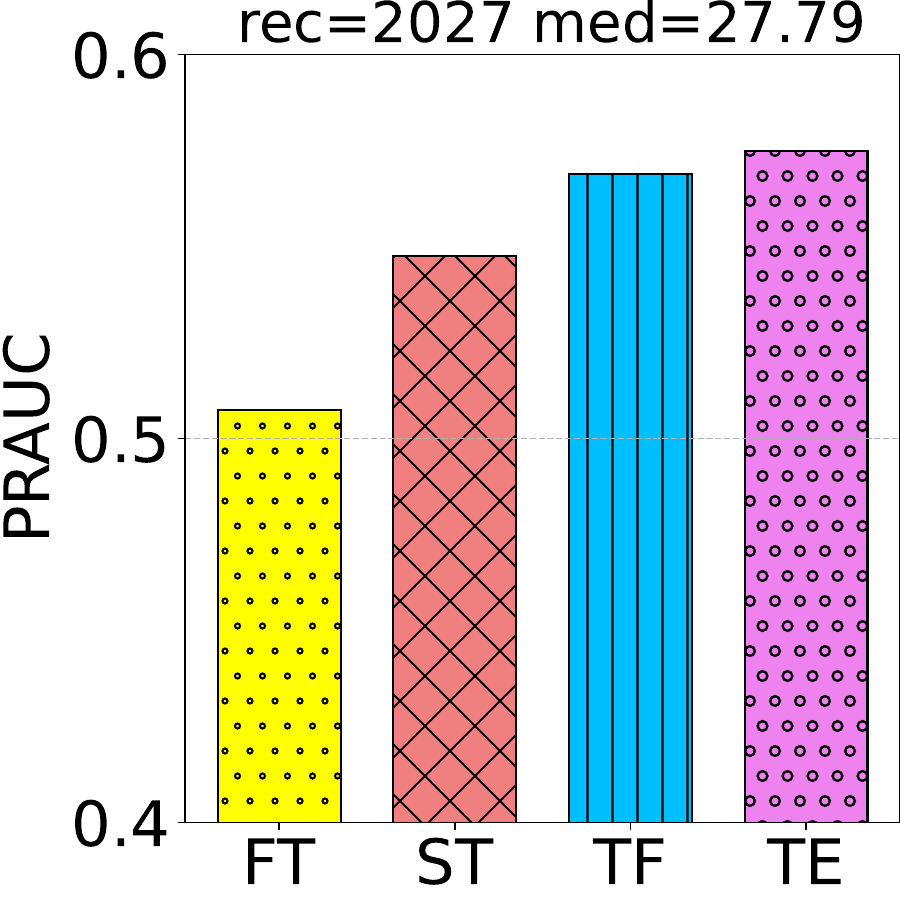}
        \caption{id=252}
    \end{subfigure}
    \end{minipage}
    \begin{minipage}[b]{0.3\linewidth}
        \centering
        \begin{subfigure}{0.95\linewidth}
        \includegraphics[scale=0.27]{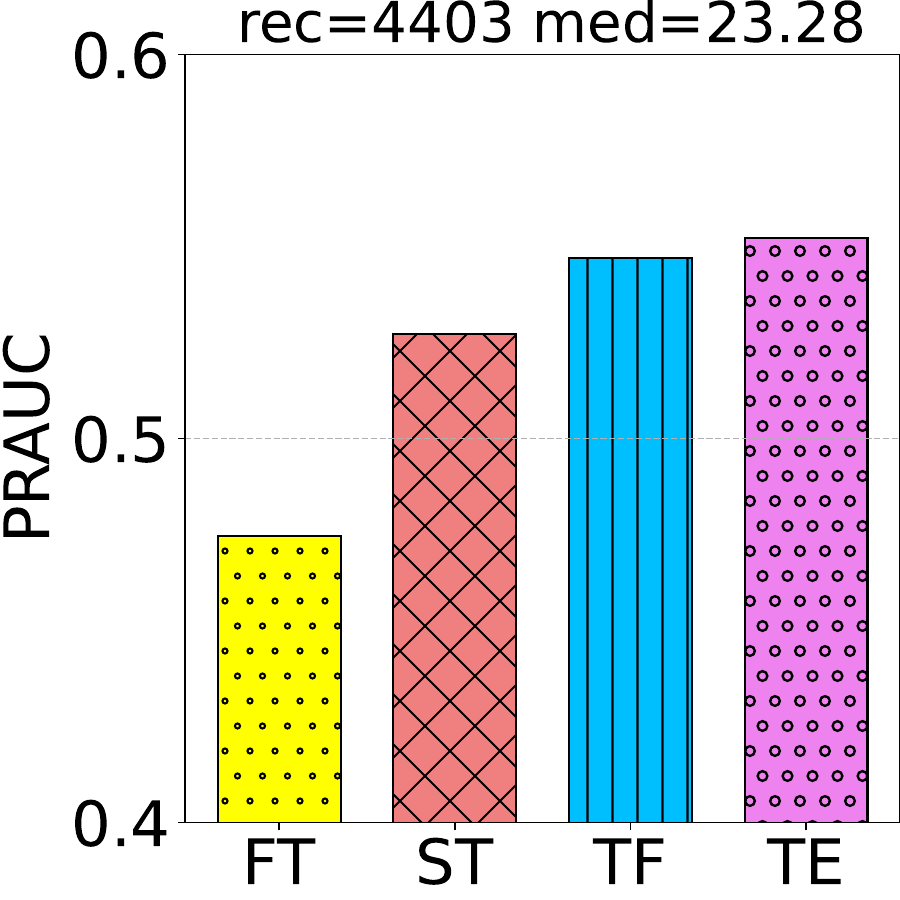}
        \caption{id=264}
    \end{subfigure}
    \end{minipage}%
    \begin{minipage}[b]{0.3\linewidth}
    \centering
    \begin{subfigure}{0.95\linewidth}
       \includegraphics[scale=0.27]{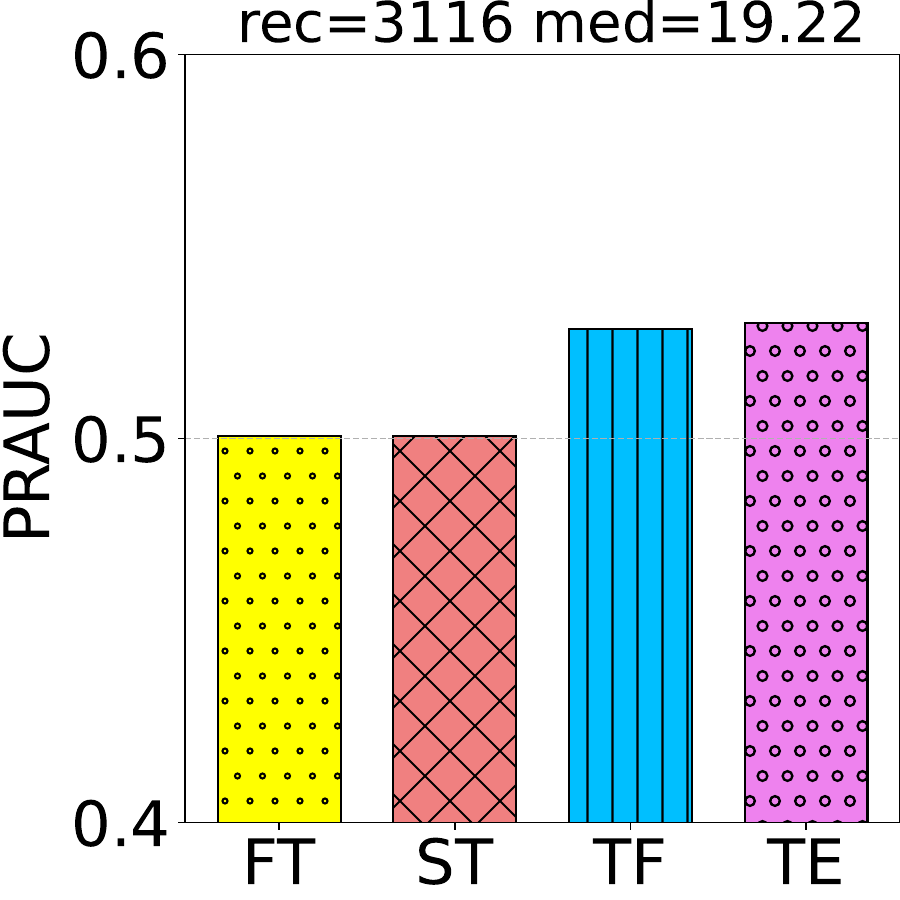}
       \caption{id=443}
    \end{subfigure}
    \end{minipage}%
    \\
\caption{The performance on randomly sampled 8 hospitals. ``rec'' represents the record number of the hospital and ``med'' is the average number of prescribed medications. FT, ST, TF and TE refer to Full-Train, Single-Train, TEMPT (finetune) and TEMPT, respectively. The id is the identifier of each hospital in the original eICU database.}
\label{appendix-case}
\end{figure} 

\subsection{Case Study} \label{sec_B1}
    In this section, we aim to analyze the performance of several hospitals in detail. As shown in Figure~\ref{appendix-case}, we randomly sample 8 hospitals, and compare the performance of the proposed TEMPT and its variants for further analysis. The results show that the proposed TEMPT often outperforms the variant models in each hospital. Besides, we find that our TEMPT and TEMPT (finetune) exceed the Single-Train much more on hospital $92$, $146$ and $181$. According to the definition in Section~\ref{sec_case}, they are all small hospitals. It can be concluded that the pretrain-finetune scheme can benefit small hospitals by learning general medical knowledge from all records. As for Full-Train, we find that its performance is unstable extremely, because it only learns the distribution of all records. If the hospital's distribution is similar to the total, it can perform well and vice versa. The TEMPT consistently outperforms its finetuning variant, which illustrates that the proposed hospital-based prompt can better capture hospital-specific information.  


\section{Related Works} \label{sec_related}

In this section, we will introduce recent research studies on the aspects of medication recommendation, multi-domain recommendation and pretrain in the recommendation.

\subsection{Medication Recommendation}

    In recent years, with the advancements in recommender systems~\cite{liu2024multimodal,li2023automlp,liu2023disentangling,liu2023linrec,lin2023autodenoise,liang2023mmmlp,liu2023multi,liu2023diffusion,zhang2024m3oe,wang2023single}, medication recommendation has attracted much attention due to its practical value~\cite{ali2023deep}. It can assist doctors by giving advice on prescriptions. For most existing works, medication recommendation is formulated as recommending a set of medications given the diagnosis and procedures of the patient. Such a process is similar to the prescription procedure in the real world. The treatments, including medications and procedures, are decided by the diagnosis of patients. For example, if one patient gets ``heart disease'', the doctor may prescribe the medication ``Angiotensin'' and ``surgery'' procedure. At the same time, ``Analgesics'' may be prescribed to assist ``surgery'', which indicates the medications also need to match the procedure. Therefore, medication recommendation models are conditioned on diagnosis and procedures. Most existing works can be categorized into two groups: instance-based and longitudinal models.

    \textit{\textbf{Instance-based Models.}} This thread of models~\cite{zhang2017leap,tan20224sdrug,zhang2022knowledge} recommend medication sets only based on the patient's current conditions. For instance, Leap~\cite{zhang2017leap} models the medication recommendation as a sequential decision-making problem and only takes the diagnosis set as input. It first captures the relationships between several diseases. Recently, Tan et al. ~\cite{tan20224sdrug} propose a symptom-based model. This model only uses the symptom and medication set and recommends medication by comparing the representations of these two sets. Besides, it considers some practical constraints when recommending drug sets, such as safety and size. MedRec~\cite{zhang2022knowledge} introduces a knowledge graph to extract the relationship between symptoms and medication sets. Our work also belongs to this category.

    \textit{\textbf{Longitudinal Models.}} Longitudinal models~\cite{le2018dual,shang2019gamenet,shang2019pre,yang2021safedrug,Yang2021ChangeMM,wang2021self,wu2022conditional,zheng2022interaction} make use of patient's historical records to get the medical information. For example, DMNC~\cite{le2018dual} adopts a memory network to model the intra-view and inter-view relations of a patient's historical diagnosis and procedures. 
    Metacare++~\cite{tan2022metacare++} adopts the meta-learning technique to capture the relationships between patients' clinical visits for cold-start diagnosis prediction.
    Shang et al.~\cite{shang2019gamenet} firstly use the graph to capture patients' longitudinal information and drug-drug interactions, which is important for drug safety. In detail, a novel dynamic memory mechanism is devised to utilize the information contained in the graph. To further decrease the drug-drug interaction rate, Yang et al.~\cite{yang2021safedrug} propose a controllable loss. Besides, for a more precise recommendation, SafeDrug also constructs the drug molecule graph and utilizes the message-passing neural network to extract the relationships between medications further. MICRON~\cite{Yang2021ChangeMM} and COGNet~\cite{wu2022conditional} consider the recommended medication set changed and copied from the patient's historical medication set. In particular, MICRON identifies the medication change between two successive visits and recommends a medication set based on the last visit. Compared to MICRON, COGNet captures the relationship between current and all historical records, and recommends by copying medications from previous prescriptions. It is worth noting that G-Bert~\cite{shang2019pre} also applies the pretrain technique to medication recommendation. In detail, it pretrains a visit encoder with single visit records and finetunes the encoder for modeling multi-visit records. Though it also proposes to pretrain for a single visit, it has two-fold differences from the proposed TEMPT: (i) G-Bert is designed for a single hospital. (ii) G-Bert is a longitudinal model, which is not for instance-based conditions especially.

    Recently, the advancements in large language models have revolutionized the research in many intelligent healthcare applications~\cite{liu2024moelora,xu2024editing,xu2024mitigating,li2023rest,li2023towards,wang2023doctor,zheng2022ddr,zheng2022cbr}, including medication recommendation. For example, Wornow~\etal~\cite{wornow2024zero} propose to construct proper prompts to motivate LLMs for clinical tasks. Besides, Liu~\etal~\cite{liu2024leader} modify the output layer of LLMs to fit medication recommendation and further propose a novel distillation method to transfer the semantics of LLMs to a small model.

    However, the previous works mentioned above aim to recommend medications for doctors from one hospital with abundant medical records, which ignores those small hospitals. To our knowledge, this paper is the first work to focus on multi-center medication recommendations.

\subsection{Multi-domain Recommendation}

     Recently, some multi-domain recommendation research studies have emerged~\cite{gao2023autotransfer,wang2023plate,li2023hamur,wang2024diff,zhang2024m3oe,gao2024hierrec,fu2023unified,wang2024llm4msr,li2022gromov,fan2021attacking,fan2023adversarial,jia2024d3}, aiming to recommend proper items for several domains simultaneously. There are two lines of work in this field, \ie multi-task learning works and multi-domain CTR prediction works. The first category regards each domain as a task and thus can learn the relationships between various domains. Among these works, MMOE~\cite{ma2018modeling} and PLE~\cite{tang2020progressive} are the two most popular multi-task learning methods. MMOE designs several expert networks to model the specific and shared information of each task, and adopts gate networks to control the contributions of experts. Compared with MMOE, PLE designs specific and shared expert networks to better model task inter-relations. The shortcoming of this line of work is that they do not fit too many domains. The other category is closely related to multi-domain CTR prediction. For example, DisenCDR~\cite{cao2022disencdr} extracts domain-specific and domain-shared user representations to share similar preferences across various domains. Besides, to alleviate the problem of entanglement of these two types of representation, it also proposes an exclusive regularizer to split them. Similar to DisenCDR, SAR-Net~\cite{shen2021sar} captures the user's cross-scenario preference by two designed attention modules and utilizes multi-expert networks to model the information of different scenarios. However, DisenCDR and SAR-Net both require users to have interactions in several domains simultaneously, which is different from the setting of multi-center medication recommendations in this paper. STAR~\cite{sheng2021one} devises the domain-specific and domain-shared feed-forward networks to recommend for users in various scenarios. Since it only inputs features of domain and item, it can be adapted to the problem in this paper.

\subsection{Pretrain in Recommendation}

    Pretrained large model has prevailed in computer vision~\cite{dosovitskiy2020image,liu2021swin} and natural language processing~\cite{devlin2018bert,sun2020ernie,brown2020language} and other fields~\cite{zhang2023promptst}, but it is under different conditions in the field of recommender system (RS). RS is founded on the non-semantic id, which hinders pretraining a universal model that can fit many scenarios. Therefore, existing pretrain works for recommendation are developed from two aspects. One aspect~\cite{hou2022towards,geng2022recommendation,cui2022m6} uses semantic information of items, such as review, description, etc., to pretrain a large semantic model. These works can benefit from the advancement in CV and NLP. For example, UniSRec~\cite{hou2022towards} proposes to learn universal item representations by the textual descriptions of each item. M6~\cite{cui2022m6} transforms the recommendation task to a pure language understanding or generation problem, and pretrain a large model that can be adapted to many downstream recommendation tasks by finetuning. However, such textual information does not exist in most recommendation situations, so the other line of work only utilizes the item's identity to be more practical. These works mainly adopt the pretrain technique to ease the data sparsity problem and get high-quality embeddings for general recommendation models. For example, S3Rec~\cite{zhou2020s3} adopts several self-supervised tasks to inject attribute and subsequence correlations into item embeddings. Hao et al.~\cite{hao2021pre} and Wu et al.~\cite{wu2021self} propose to pretrain the graph-based recommendation model to enhance the long-tail and cold-start embeddings, respectively. Our work belongs to the latter aspect, which only uses the identity of medication.

    Prompt, as an efficient method, is widely used to tune the large pretrained model~\cite{liu2021pre}. Li et al.~\cite{li2021prefix} propose a continuous prefix prompt to tune the GPT-2~\cite{radford2019language} for some NLP tasks. Then, VPT~\cite{jia2022visual} shows that the idea of continuous prompt also works for large vision pretrained models. It sets different learnable prompt vectors when tuning for various downstream vision tasks. Recently, some researchers have also adopted the prompt to recommender systems, such as PEPLER~\cite{li2022personalized} and P5~\cite{geng2022recommendation}. Similar to M6, P5~\cite{geng2022recommendation} first transfers all data of recommendation to a natural language format and then pretrains a unified model. Based on the pretrained model, it proposes an instruction-based prompt to facilitate the power of zero-shot recommendation. However, P5 is a natural language-based model, which has many limitations in the general recommendation field. PEPLER~\cite{li2022personalized} is the first to explore the prompt for the typical recommendation. PEPLER devises a prompt generator based on the user's profile and only tunes it after pretraining a general sequential recommendation model. Different from PEPLER, the proposed prompt in TEMPT is devised based on various hospitals and aims at benefiting the recommendation for multi-center. As far as we know, our work is one of the pioneers to adopt prompt to a pure id-based recommendation model, which gives light to this field.

\section{Conclusion} \label{sec_conclusion}

In this paper, we propose a contrastive pretrain model with prompt tuning (TEPMT) for a more practical scenario, \ie multi-center medication recommendation. Specifically, we first design a transformer-based medical encoder to extract medical information from the input diagnosis and procedure sets. Then, the TEMPT is equipped with two self-supervised tasks to learn general medical knowledge, \ie mask prediction task and contrastive task for intra- and inter-relationships, respectively. Besides, we design a novel hospital-based prompt tuning to avoid the catastrophic forgetting problem of general finetuning and better fit each hospital's various distributions. To validate our model, we conduct extensive experiments on a multi-center medical dataset, \ie eICU. The experimental results show that the proposed model outperforms existing medication recommendation models and multi-domain recommendation models. Primarily, we find that the proposed TEMPT is also beneficial for the hospitals on a small scale compared with all baseline models, which is important to the healthcare system. Furthermore, since we only need to train a small proportion of parameters while prompt tuning, the storage and computation costs are much smaller than the general finetuning strategy. 
Despite the superior performance in multi-center medication recommendation, one possible limitation is that the proposed TEMPT does not take drug-drug interaction into consideration. We leave it to the future work.
Besides, due to the privacy requirements of patients and hospitals, the medical data from multi-center cannot be accessed easily in the real world. Therefore, in the future, we will focus on combining federated learning with the proposed pretrain model. 
Furthermore, most existing research studies only adopt the diagnosis and procedure identity to model the collaborative information, while ignoring patient's personal profiles, such as allergies and age, and semantic relationships between diagnosis, procedure and medications. Thus, exploring how to utilize those informative features by a powerful large language model for medication recommendation is also an interesting and promising direction.

\section{Acknowledgements}
    This research was partially supported by 
    Research Impact Fund (No.R1015-23), APRC - CityU New Research Initiatives (No.9610565, Start-up Grant for New Faculty of CityU), CityU - HKIDS Early Career Research Grant (No.9360163), Hong Kong ITC Innovation and Technology Fund Midstream Research Programme for Universities Project (No.ITS/034/22MS), Hong Kong Environmental and Conservation Fund (No. 88/2022), and SIRG - CityU Strategic Interdisciplinary Research Grant (No.7020046), Huawei (Huawei Innovation Research Program), Tencent (CCF-Tencent Open Fund, Tencent Rhino-Bird Focused Research Program), Ant Group (CCF-Ant Research Fund, Ant Group Research Fund), Alibaba (CCF-Alimama Tech Kangaroo Fund No. 2024002), CCF-BaiChuan-Ebtech Foundation Model Fund, and Kuaishou,  
    National Natural Science Foundation of China (No.62192781, No.62177038, No.62293551, No.62277042, No.62137002, No.61721002, No.61937001, No.62377038), Project of China Knowledge Centre for Engineering Science and Technology, ``LENOVO-XJTU'' Intelligent Industry Joint Laboratory Project. 

\bibliographystyle{ACM-Reference-Format}
\bibliography{main}


\end{document}